\documentclass[11pt]{article}

 \csname
@addtoreset\endcsname{equation}{section}

\def\a{\alpha}  

\def\b{\beta}   
\def\c{\gamma} 
\def\C{\Gamma}
\def\d{\delta} 
\def\D{\Delta}
\def\e{\epsilon} 

\def\tf{\tilde{\phi}}
\def\F{\Phi}
\def\vf{\varphi}
\def\k{\kappa}
\def\l{\lambda}
\def\L{\Lambda}
\def\m{\mu}
\def\n{\nu}
\def\r{\rho}
\def\s{\sigma}
\def\S{\Sigma}
\def\t{\tau}
\def\th{\theta}

\def\x{\xi}
\def\y{\eta}

\def\O{\Omega}

\def\cA{{\cal A}}
\def\cR{{\cal R}}
\def\cS{{\cal S}}
\def\cP{{\cal G}}

\def\cI{{\cal I}}

\def\cA{{\cal A}}

\def\yb{{\bar y}}

\def\ra{\rightarrow}

\let\la=\label
\let\bm=\bibitem
\def\nn{\nonumber}
\newcommand{\eq}[1]{(\ref{#1})}
\newcommand{\ns}[1]{{\normalsize #1}}

\newcommand{\w}[1]{\\[0.#1cm]}
\def\eqs#1#2{(\ref{#1}-\ref{#2})}

\def\be{\begin{equation}}
\def\ee{\end{equation}}
\def\bea{\begin{eqnarray}}
\def\eea{\end{eqnarray}}
\def\ba{\begin{array}}
\def\ea{\end{array}}

\def\ft#1#2{{\textstyle{{\scriptstyle #1}
\over {\scriptstyle #2}}}} \def\fft#1#2{{#1 \over #2}}

\def\ket#1{|#1\rangle}

\def\ta{
\begin{table}[t]
\begin{center}
{\footnotesize \tabcolsep=1mm
\begin{tabular}{|c|cccccccc|}\hline
& & & & & & & &  \\
{\large${}_{|Z|}\backslash s$} & $0$ & \ns{$\ft12$} & $1$ &
\ns{$\ft32$} & $2$ & \ns{$\ft52$} &
$3$ & $\cdots$ \\
& & & & & & & &   \\ \hline
& & & & & & & &   \\
$0$ & $6$ & $4$ & $1$ & & & & & \\
$1$ & $1$ & $4$ & $6$ & $4$ & $1$ & & & \\
$2$ & & & $1$ & $4$ & $6$ & $4$ & $1$  & \\
$3$ & & & & & $1$ & $4$ & $6$ & $\cdots$ \\
$4$ & & & & & & & $1$ & $\cdots$ \\
$\vdots$ & & & & & & & & \\ \hline
\end{tabular}}
\end{center}
\caption{{\footnotesize The superdoubletons with integer $Z$. The
spin is defined as $s=j_L+j_R$ and the entries denote the $SU(4)$
representations. Suitable reality conditions need to be imposed.
The $U(1)_Y$ charges of $1$, $4$ and $6$ are $0$, $\pm1$ and
$\pm2$, respectively. For $|Z|\geq 1$ each supermultiplet consists
of a superdoubleton and its CPT-conjugate (with opposite signs of
$Z$) combined together. The $Z=0$ multiplet is the
(self-conjugate) $d=4,N=4$ Yang-Mills supermultiplet. }} \la{ta}
\end{table}
}

\def\tb{
\begin{table}[t]
\begin{center}
{\footnotesize \tabcolsep=1mm
\begin{tabular}{|c|ccccccccc|}\hline
& & & & & & & & & \\
{\large${}_{2|Z|}\backslash s$} & $0$ & \ns{$\ft12$} & $1$ &
\ns{$\ft32$} & $2$ & \ns{$\ft52$} &
$3$ & \ns{$\ft72$} & $\cdots$ \\
& & & & & & & & & \\ \hline
& & & & & & & & & \\
$1$ & $4$ & $6+1$ & $4$ & $1$ & & & & & \\
$3$ & & $1$ & $4$ & $6$ & $4$ & $1$ & & & \\
$5$ & & & & $1$ & $4$ & $6$ & $4$ & $1$  & \\
$7$ & & & & & & $1$ & $4$ & $6$ & $\cdots$ \\
$9$ & & & & & & & & $1$ & $\cdots$ \\
$\vdots$ & & & & & & & & & \\ \hline
\end{tabular}}
\end{center}
\caption{{\footnotesize The superdoubletons with half-integer
$|Z|$; see caption of Table 1. Here suitable reality as well as
self-duality conditions need to be imposed. }} \la{tb}
\end{table}
}

\def\tc{
\begin{table}[t]
\begin{center}
{\footnotesize \tabcolsep=1mm
\begin{tabular}{|c|cc|cccccccccccc|}\hline
& & & & & & & & & & & & & & \\
{\large${}_{\ell}\backslash s$} & $0$ & \ns{$\ft12$} & $1$ &
\ns{$\ft32$} & $2$ & \ns{$\ft52$} &
$3$ & \ns{$\ft72$} & $4$ & $\ft92$ & $5$ & $\ft{11}2$ & $6$ & $\cdots$ \\
& & & & & & & & & & & & & & \\ \hline
& & & & & & & & & & & & & & \\
$0$ & $42$ & $48$ & $27$ & $8$ & $1$ &
& & & & & & & & \\
$1$ & $1$ & $8$ & $28$ & $56$ & $70$ & $56$ & $28$ & $8$ & $1$ &
& & & &  \\
$2$ & & & & & $1$ & $8$ & $28$ & $56$ & $70$
& $56$ & $28$ & $8$ & $1$  &  \\
$3$ & & & & & & & & & $1$ & $8$ & $28$ & $56$ & $70$
& $\cdots$ \\
$4$ & & & & & & & & & & & & & $1$ & $\cdots$ \\
$\vdots$ & & & & & & & & & & & & & &  \\ \hline
\end{tabular}}
\end{center}
\caption{{\footnotesize The symmetric tensor product of two $N=4$
SYM doubletons arranged into levels $\ell=0,1,2...$ of $N=8$
$AdS_5$ superalgebra multiplets. Each level appears
in the product once and only once, and consists of $USp(8)$
representations, some of which are reducible: $28=27+1$,
$56=48+8$, $70=42+27+1$. Under $SU(4)\times U(1)_Y $:
$8=4_1+\bar{4}_{-1}$, $27=15_0+6_{2}+\bar{6}_{-2}$,
$42=20'_0+10_2+\bar{10}_{-2}+1_4+\bar{1}_{-4}$,
$48=20_1+\bar{20}_{-1}+4_3+\bar{4}_{-3}$. The $U(1)_Y$ charges are
determined by $j_L-j_R+\ft12 Y=0$. The level $0$ multiplet
is the usual $D=5,N=8$ supergravity multiplet. The states in the $s<1$ sector arise as the physical states in the
master scalar field $\Phi$, as shown in Table \ref{tf}. For $s\geq 1$, the
states with $Y=0,\pm 1$ arise as physical states in the gauge
fields corresponding to the generators of the higher spin algebra
listed in Table \ref{td}. Those with $Y=\pm 2, \pm 3, \pm 4$
arise as physical states in the master scalar field $\Phi$. With
the exception noted in Table \ref{tf}, these have dual gauge
fields corresponding to the generators listed in Table \ref{te}.
}} \la{tc}
\end{table}
}

\def\td{
\begin{table}[t]
\begin{center}
{\footnotesize \tabcolsep=1mm
\begin{tabular}{|c|cccccccccccc|}\hline
& & & & & & & & & & & & \\
{\large${}_{\ell}\backslash s$} & $1$ & \ns{$\ft32$} & $2$ &
\ns{$\ft52$} &
$3$ & \ns{$\ft72$} & $4$ & $\ft92$ & $5$ & $\ft{11}2$ & $6$ & $\cdots$ \\
& & & & & & & & & & & & \\ \hline
& & & & & & & & & & & & \\
$0$ & $15$ & $4$ & $1$ & & & & & & & & & \\
$1$ & $16'$ & $24$ & $36$ & $24$ & $16'$ & $4$ & $1$ & & & & &  \\
$2$ & & & $1$ & $4$ & $16'$ & $24$ & $36$
& $24$ & $16'$ & $4$ & $1$  &  \\
$3$ & & & & & & & $1$ & $4$ & $16'$ & $24$ & $36$ & $\cdots$ \\
$4$ & & & & & & & & & & & $1$ & $\cdots$ \\
$\vdots$ & & & & & & & & & & & &  \\ \hline
\end{tabular}}
\end{center}
\caption{{\footnotesize The $hs(2,2|4)$ generators with $Y=0,\pm
1$. The entries are $SU(4)\times U(1)_Y$ representations as
follows: $16'=15_0+1_0$, $24=20_1+4_1$, $36=20'_0+15_0+1_0$.}
These generators are associated with the physical gauge fields all
of which have their associated Weyl tensors. The spin $s$ defined
by $s=1+\ft12(n_y+n_\yb)$ is that of the gauge field associated
with the generator. The level $\ell$ is defined by
$\ell=\ft14(n_y+n_\yb+n_\th+n_{\bar{\th}}-2)$. } \la{td}
\end{table}
}

\def\te{
\begin{table}[t]
\begin{center}
{\footnotesize \tabcolsep=1mm
\begin{tabular}{|c|cccccccccc|}\hline
& & & & & & & & & & \\
{\large${}_{\ell}\backslash s$} & $2$ & \ns{$\ft52$} &
$3$ & \ns{$\ft72$} & $4$ & $\ft92$ & $5$ & $\ft{11}2$ & $6$ & $\cdots$ \\
& & & & & & & & & & \\ \hline
& & & & & & & & & & \\
$1$ & $16$ & $4$ & $6$ & & & & & & &  \\
$2$ & & & $6$ & $4$ & $16+1$ & $4$ & $6$ & & &  \\
$3$ & & & & & & & $6$ & $4$ & $16+1$ & $\cdots$ \\
$\vdots$ & & & & & & & & & &  \\ \hline
\end{tabular}}
\end{center}
\caption{{\footnotesize The $hs(2,2|4)$ generators with
$Y=\pm2,\pm3,\pm4$. The entries are $SU(4)\times U(1)_Y$
representations as follows: $16=10_2+6_2$, $4_3$ and $1_4$. These
generators are associated with gauge fields dual to generalized
anti-symmetric tensor fields contained in the scalar master field
$\Phi$; see Table \ref{tf} for $s\geq 1$. Further notation is
defined in Table \ref{td}.}} \la{te}
\end{table}
}

\def\tf{
\begin{table}[t]
\begin{center}
{\footnotesize \tabcolsep=1mm
\begin{tabular}{|c|cc|cccccccccccc|}\hline
& & & & & & & & & & & & & & \\
{\large${}_{\ell}\backslash s$} & $0$ & \ns{$\ft12$} & $1$ &
\ns{$\ft32$} & $2$ & \ns{$\ft52$} &
$3$ & \ns{$\ft72$} & $4$ & $\ft92$ & $5$ & $\ft{11}2$ & $6$ & $\cdots$ \\
& & & & & & & & & & & & & & \\ \hline
& & & & & & & & & & & & & & \\
$0$ & $42$ & $48$ & $\underline{6}$ & & & & & & & & & & & \\
$1$ & $1$ & $8$ & $\underline{6}$ & $\underline{4}$ & $16+\underline{1}$ & $4$ & $6$ & & & & & & & \\
$2$ & & & & & & & $6$ & $4$ & $16+1$ & $4$ & $6$ & & & \\
$3$ & & & & & & & & & & & $6$ & $4$ & $16+1$ & $\cdots$ \\
$\vdots$ & & & & & & & & & & & & & &  \\ \hline
\end{tabular}}
\end{center}
\caption{{\footnotesize The physical matter fields contained in
the master scalar field. The entries are the following
$SU(4)\times U(1)_Y$ representations for $s<1$:
$42=20'_0+10_2+\bar{10}_{-2}+1_4+\bar{1}_{-4}$,
$48=20_1+\bar{20}_{-1}+4_3+\bar{4}_{-3}$, $8=4_1+\bar{4}_{-1}$
and $1_0$; for $s\geq 1$: $6_2$, $4_3$, $16=10_2+6_2$ and $1_4$.
The spin $s\geq 1$ sector is realized in the field theory in
terms of generalizations of the anti-symmetric two-form
potential. These fields obey self-duality in $D=5$ and have dual
one-form gauge fields corresponding to the generators given in
Table \ref{te}, with the exception of the underlined
representations, which have no one-form duals. Here the form
degree refers to the number of curved indices as opposed to the
tangential multi-spinor indices arising from the
$(y,\yb)$-expansion.}} \la{tf}
\end{table}
}

\newcommand{\hoch}[1]{$\, ^{#1}$}

\newcommand{\tamphys}{\it\small Center for Theoretical Physics, Texas
A\&M University, College Station, TX 77843, USA}

\newcommand{\groningen}{\it\small Institute for Theoretical Physics,
University of Groningen, 9747 AG Groningen,The Netherlands}

\newcommand{\auth}{\large E. Sezgin\hoch{\dagger}and P. Sundell\hoch{\star} }

\begin{document}

\hfill{CTP-TAMU-25/01}

\hfill{UG-01-31}

\hfill{hep-th/0107186}


\vspace{20pt}

\begin{center}


{\Large \bf Towards Massless Higher Spin Extension } \w3
{\Large\bf of D=5, N=8 Gauged Supergravity}


\vspace{20pt}

\auth

\vspace{15pt}

\begin{itemize}

\item[$^\dagger$] \tamphys

\item[$^\star$] \groningen

\end{itemize}

\vspace{30pt}

{\bf Abstract}

\end{center}

The $AdS_5$ superalgebra $PSU(2,2|4)$ has an infinite dimensional
extension, which we denote by $hs(2,2|4)$. We show that the
gauging of $hs(2,2|4)$ gives rise to a spectrum of physical
massless fields which coincides with the symmetric tensor product
of two $AdS_5$ spin-1 doubletons (i.e. the N=4 SYM multiplets
living on the boundary of $AdS_5$). This product decomposes into
levels $\ell=0,1,2,..,\infty$ of massless supermultiplets of
$PSU(2,2|4)$. In particular, the $D=5, N=8$ supergravity
multiplet arises at level $\ell=0$. In addition to a master gauge
field, we construct a master scalar field containing the
$s=0,1/2$ fields, the anti-symmetric tensor field of the gauged
supergravity and its higher spin analogs. We define the
linearized constraints and obtain the linearized field equations
of the full spectrum, including those of $D=5,N=8$ gauged
supergravity and in particular the self-duality equations for the
2-form potentials of the gauged supergravity (forming a 6-plet of
$SU(4)$), and their higher spin cousins with $s=2,3,...,\infty$.

\pagebreak

\setcounter{page}{1}


\section{Introduction}


Higher spin gauge theories, for sometime considered in their own
right (see, for example, \cite{us} for a review), are likely to
find their niche in M-theory, where it is natural to expect
similar structures in the limit of high energies. In fact, the
first indication of a possible connection between higher spin
gauge theory and the physics of extended objects was pointed out
long ago \cite{bsst} in the context of the eleven dimensional
supermembrane on $AdS_4\times S^7$. More recently, tensionless
type IIB closed string theory in a background with non-zero
three-brane charge has been argued to be described by a higher
spin gauge theory expanded around $AdS_5$ \cite{su1,su2}. The
tensionless limit requires vanishing string coupling. The higher
spin gauge theory therefore describes stringy interactions of a
new kind. In the limit of large three-brane charge, i.e. weak
curvature, the five-dimensional Planck scale is much larger than
the inverse $AdS$-radius, and the higher spin gauge theory has an
effective field theory description in terms of a curvature
expansion valid at energies much smaller than the Planck scale.

Anti-de Sitter spacetime arises naturally in higher spin gauge
theory. This suggests that higher spin gauge theory may play a
role in understanding the strong version of the Maldacena
conjecture, i.e. not relying on taking the low-energy limit. In
this context, tensionless closed strings would be dual to
tensionless open strings with vanishing t' Hooft coupling. A
starting point is to examine currents formed out of free
superdoubletons in 4d Minkowski
space.

The superdoubleton representations in question are the ultra
short representations of $AdS_5$ superalgebra $PSU(2,2|4)$ which
have a fixed radial dependence in $AdS_5$ and thus they live on
the boundary of $AdS_5$ \cite{g1}. A group theoretically precise
definition will be given later but for the present discussion let
us note that these are massless representations of the $4d$ super
Poincar\'e algebra which admit the realization of the $4d$
superconformal symmetry, which is isomorphic to the $AdS_5$
supersymmetry. Moreover, there are infinitely many such $AdS_5$
doubleton supermultiplets which are listed in Tables 1 and 2.
Note that the shortest one is the (self-conjugate) Yang-Mills
supermultiplet, and that the usual $4d, N=4$ supergravity
multiplet also figures in the list (the level one supermultiplet
in Table 1). In fact, the tensor product of any pair of
superdoubletons decomposes into an infinite set of $AdS_5$
massless supermultiplets. We will focus our attention on the
higher spin gauge theory based on the $N=4 SYM$ doubletons but we
shall come back to this point in the the final section.

Assuming the strong version of the Maldacena conjecture, the
boundary currents mentioned above should be formed strictly out
of the Yang-Mills multiplet (with global $SU(N)$ symmetry).
The symmetric tensor product of two
such multiplets decomposes into an infinite set of massless
supermultiplets of the $AdS_5$ superalgebra \cite{g1,g2,fz} which
arrange themselves into levels $\ell=0,1,2,...,\infty$ which
contain the $D=5, N=8$ gauged supergravity multiplet at level
$\ell=0$. This suggests that all the conserved boundary currents
made out of two $AdS_5$ Yang-Mills doubletons (in the $g_{YM} \ra
0$ limit) couple naturally to the massless higher spins of the
higher spin $D=5, N=8$ supergravity theory considered here. This
in itself, of course, is not enough of an evidence for a
connection with string theory. See, however, \cite{su1,su2},
where further arguments are given in favor of such a connection.

In trying to make contact with string theory, an important issue
is whether open string interactions survive the tensionless limit,
and in that case, whether these can be accommodated in the
superdoubleton theory, perhaps in the form of higher derivative
deformations. These interactions would be different in nature
from the ordinary Yang-Mills interactions, which preserve
conformal invariance but break higher spin symmetries
\cite{anselmi}.

Motivated by these considerations, we recently constructed a
bosonic higher spin algebra extension of the $AdS_5$ group based
on spin zero doubletons, and gave its linearized gauge theory in
five dimensions \cite{us}. In this paper we generalize this
result to an extension of the the $AdS_5$ superalgebra
$PSU(2,2|4)$, which we shall denote by $hs(2,2|4)$, based on the
Yang-Mills superdoubleton. The main results of this paper are:

\begin{itemize}

\item The identification of the symmetry group of the higher spin gauge
theory as the higher spin extension of the $AdS_5$ superalgebra
$PSU(2,2|4)$ in which an ideal generated by a central element is
modded out \cite{5dv1}.

\item The definition of the massless spectrum as the product of
two $AdS_5$ Yang-Mills superdoubletons.

\item The identification of the spectrum with the physical
field content of one-form and zero-form master fields subject to
(linearized) constraints.

\item The linearized field equations of the full spectrum, including those
of gauged $D=5,N=8$ supergravity \cite{5d1,5d2} and in particular
the self-duality equations for the 2-form potentials of the gauged
supergravity (forming a 6-plet of $SU(4)$), and their higher spin
cousins with $s=2,3,...,\infty$.

\end{itemize}

The analysis in this paper is linearized. However, the field
content is complete in the sense that the full spectrum of the
higher spin algebra is included, as required by unitarity and
ultimately by the consistency of the full interacting theory. In a
recent paper \cite{5dv2}, certain cubic interactions of a bosonic
higher spin theory in five dimensions were constructed using an
action. In particular, these interactions do not involve the
matter fields which are essential for the description of the full
interacting theory. It is clearly desirable to construct the full
interactions of the $D=5, N=8$ higher spin gauge theory, which we
believe should exist. We are currently studying this problem.

The paper is organized as follows. In Section 2, we give the
necessary details of the representation theory of $SU(2,2|4)$,
with particular emphasis on the oscillator realization. In
Section 3, the higher spin extension $hs(2,2|4)$ of $PSU(2,2|4)$
is defined. In Section 4, the spectrum of states as the symmetric
product of two $N=4$ SYM doubletons is defined (the decomposition
of the spectrum under $PSU(2,2|4)\times U(1)_Y$ is shown in
Appendix A). In Section 5, the kinematics of the field theoretical
realization of $hs(2,2|4)$ as a gauge theory in five-dimensional
spacetime is described. In Section 6, the field content of the
master scalar field is determined. In Section 7, the linearized
constraints and the proof of their integrability are given. In
Section 8, the resulting field equations, including self-duality
equations for higher spin fields generalizing those of the $D=5$,
$N=8$ gauged supergravity theory, are found (the
details of harmonic analysis used to compute lowest energies are
provided in Appendix B). In Section 9, we comment further on our
results and suggest future directions.


\section{Doubletons and Massless Irreps of $PSU(2,2|4)$}


The generators of $SU(2,2|4)$ (and also its higher spin extension)
can be constructed in terms of the bosonic oscillators $y_\a$
($\a=1,..,4$), which are $SO(4,1)$ Dirac spinors, and the
fermionic oscillators $\theta_i$ ($i=1,...,4$), which are in the
fundamental representation of $SU(4)$. The oscillator algebra
reads

\be
y_\a\star \bar{y}^\b-\bar{y}^\b\star y_\a=2\d_\a^\b\ ,\quad
\bar{\th}^i\star\th_j+\th_j\star\bar{\th}^i=2\d^{i}_j\ ,
\la{osc}
\ee
where $\star$ denotes the operator product. The remaining
(anti)commutators vanish. We also define a Weyl ordered product as
follows:

$$
y_\a\bar{y}^\b=y_\a \star \bar{y}^\b-\d_\a^\b\ ,\quad
\bar{y}^\a y_\b=\bar{y}^\a \star y_\b+\d_\b^\a\ ,
$$
\be
\bar{\th}^i\th_j=\bar{\th}^i\star\th_j-\d^i_j\,\quad
\th_i\bar{\th}^j=\th_i\star \bar{\th}^j-\d_i^j\ .
\la{weyl}
\ee
The Weyl ordered product extends in a straightforward fashion to
arbitrary polynomials of the oscillators; see \cite{us} for
details and the $SO(4,1)$ spinor conventions. The generators of
$SU(2,2|4)$ consist of the bosonic $SU(2,2)\times SU(4)\times
U(1)_Z$ generators denoted by $M^\a{}_\b$, $T^i{}_j$, $Z$,
respectively, and the supersymmetry generators $Q_\a^i$. They can
be realized as

\bea
M^\a{}_\b &=& \ft12 \bar{y}^\a y_\b -\ft14 \d^a_b K\ , \nn\w2
T^i{}_j &=& \ft12 \bar{\th}^\a \th_j  -\ft14 \d^i_j X\ ,
\nn\w2
Z&=& \ft12(K+X)\ , \nn\w2 Q_\a^i &=&\bar{\theta}^i y_\a\ ,
\eea
where
\be
K \equiv \ft12 \bar{y} y\ ,\quad\quad X\equiv \ft12 \bar{\th}\th\ .
\ee
In tensorial basis the $SU(2,2)$ generators are
$M_{AB}=(\S_{AB})_\a{}^\b M^\a{}_\b\ (A=0,1,2,3,5,6)$ where
$\S_{AB}$ are the van der Waerden symbols of $SO(4,2)$. The AdS
energy generator is $E=M_{60}$. The generator $Z$ commutes with
all the generators and therefore acts like a central charge. By
factoring out this Abelian ideal one obtains the simple Lie
superalgebra $PSU(2,2|4)$. As we shall see later, the higher spin
algebra we will work with is a natural extension of $PSU(2,2|4)$.
There also exists an outer automorphism group $U(1)_Y$ generated
by $X$. The $U(1)_Y$ charge is denoted by $Y$. In summary, the
$PSU(2,2|4)\times U(1)_Z \times U(1)_Y$ algebra takes the
schematic form $[\L,\L]= \L+Z$, $[Z,\L]=0$ and $[Y,\L]=\L$, where
$\L$ are the generators of $PSU(2,2|4)$ and $Z$ appears only in
the anti-commutator of two conjugate supersymmetry generators.
With $Z$ modded out, the $PSU(2,2|4)$ algebra closes and $Y$ acts
as an outer automorphism. In fact, this structure will generalize
to the higher spin algebra.

Representations of $PSU(2,2|4)$ are representations of
$SU(2,2|4)$ with vanishing central charge $Z$. Physical
representations of $SU(2,2|4)$ consist of a multiplet of lowest
weight representations $[D(j_L,j_R;E_0)\otimes R \otimes Z]_Y$ of
the bosonic subalgebra $B=SU(2,2)\times SU(4)\times U(1)_Z$,
which are characterized by ground states labeled by the energy
$E_0$, the $SU(2)_L\times SU(2)_R \subset SU(2,2)$ spins
$(j_L,j_R)$, an $SU(4)$ irrep $R$, a central charge $Z$ and an
$U(1)_Y$ charge $Y$ \cite{g1}. A subset of them, forming the
Clifford vacuum, is also the ground state of $SU(2,2|4)$. The
requirement that it is also annihilated by fermionic energy
lowering operators puts restriction on its labels. Action by the
fermionic energy raising operators then generate a supermultiplet.

The oscillator realization gives rise to unitary supermultiplets;
see Appendix A. The Fock space of a single set of oscillators
decomposes into the doubleton supermultiplets listed in Table
\ref{ta} and Table \ref{tb}. These have either $E_0=j_L+1$, $j_R=0$ or
$E_0=j_R+1$, $j_L=0$. The central charge is given by
$Z=j_L-j_R+\ft12 Y$. Their $SU(4)\times U(1)_Y$ content is given
in the tables. The decomposition of the oscillator Fock space
into superdoubletons is such that each allowed value of $Z$
(integer and half-integer) occurs once and only once.

By considering tensor products of oscillator Fock spaces, one can
construct massless and massive supermultiplets. The two-fold
tensor product gives rise to massless representations \cite{g1}.
In particular, the product of a superdoubleton and its
CPT-conjugate gives rise to vanishing central charge and energies obeying
\footnote{More
generally, there are `novel' multiplets \cite{g2} that have energy
$E_0>j_L+j_R+2$.}

\be
E_0 = \left\{\ba{ll} j_L+j_R+2& \mbox{for }j_L+j_R\geq 1\\
2+\ft{|Y|}2&\mbox{for }j_L+j_R=0, \ft12 \ea\right.
\la{e0}
\ee
The $U(1)_Y$ charges are determined by
$j_L-j_R+\ft12 Y=0$. In particular, the symmetric product of two
Yang-Mills supermultiplets, which are CPT self-conjugate
superdoubletons with $Z=0$ and play an important role in the
construction of our higher spin theory, decomposes into massless
supermultiplets as shown in Table \ref{tc}.

\ta

\tb

\tc


\section{The Higher Spin Superalgebra $hs(2,2|4)$}


The definition of the higher spin extension $hs(2,2)$ of
$SU(2,2)$, with spectrum given by the product of two (bosonic)
scalar doubletons, was first given in \cite{us}. An important
feature is the modding out of an Abelian ideal, which is generated
by $K$ in the bosonic case. In this section we define a higher
spin extension $hs(2,2|4)$ of $PSU(2,2|4)$ (denoted by
$ho_0(8,8|8)$ in \cite{5dv1}) by the coset $\cP/\cI$, where $\cP$
is a Lie supersubalgebra of the algebra $\cA$ of arbitrary
polynomials of the oscillators in \eq{osc} and $\cI$ is an ideal
of $\cP$ generated by the central element $Z$. We also define the
physical spectrum $\cS$ of the five-dimensional higher spin gauge
theory based on $hs(2,2|4)$. The basic requirement on $\cS$ is
that it must consist of massless supermultiplets of the $N=8$
AdS$_5$ superalgebra and carry a unitary (irreducible)
representation of $hs(2,2|4)$. Given the algebra $hs(2,2|4)$ and
its massless spectrum $\cS$, we will consider the higher spin
gauge theory based on this data in the next section.

We first define a set of linear maps $\tau_\y$, labeled by a
unimodular complex parameter $\y$, acting on a Weyl ordered
function $F$ of oscillators as follows:

\be
\t_\y (F(y_\a,\yb^\a,\th_i,\bar{\th}^i))= F(\y
y_\a,-\bar{\y}\yb^\a,\y\th_i, -\bar{\y}\bar{\th}^i)\ ,\quad |\y|=1\ .
\la{tau}
\ee
These maps act as anti-involutions of $\cA$:

\be
\t_\y (F\star G)=(-1)^{FG}\t_\y(G)\star \t_\y(F)\ .
\la{ai}
\ee
The Lie superalgebra $\cP$ is defined to be the subspace of $\cA$
consisting of Grassmann even elements $P$ obeying
\footnote{The algebra $\cP$ can be enlarged by restricting $\y$
such that $\y^M=1$ for some integer $M$. For example, taking
$\y=i$ one obtains a higher spin algebra, denoted by $ho(8,8|8)$
\cite{5dv1}, with the finite subalgebra $OSp(8|8)\supset
SO(8)\times Sp(8,R)$, where $SO(8)\supset SU(4)$ and
$Sp(8,R)\supset SU(2,2)$.}

\be
\cP\ :\quad  \t_\y(P)=-P\ ,\quad (P)^{\dagger}=-P\ ,
\la{tco}
\ee
and with Lie bracket

\be [P,Q]=P\star Q-Q\star P\ .\la{l}\ee
We decompose $\cP$ into levels labeled by $\ell=0,1,2,...$, such
that the $\ell$th level is given by all elements of the
form\footnote{One can equivalently choose a basis in which the
elements are traceless in the superindex $R=(\a,i)$. The two bases
are related by redefinitions involving finite linear
combinations.}:

\be P^{(k)}(m,n;p,q) = {1\over m!n!p!q!} Z^{*k}\star\left[
P^{(k)}{}^{\b_1\dots \b_n}_{\a_1\dots \a_m}{}^{j_1\dots
j_q}_{i_1\dots i_p} \yb^{\a_1}\cdots \yb^{\a_m}y_{\b_1}\cdots
y_{\b_n}\bar{\th}^{i_1}\cdots \bar{\th}^{i_p}\th_{j_1}\cdots
\th_{j_q}\right] \ ,\la{alge1}\ee

$$m+n+p+q=4\ell+2\ ,\quad m+p=n+q\ ,\quad n+q=1\mbox{ mod }2\
,$$

\be P^{(k)}{}^{\c\b_1\dots \b_{n-1}}_{\c\a_1\dots
\a_{m-1}}{}^{j_1\dots j_q}_{i_1\dots i_p} =0 \ ,\la{alge2}\ee
where we use the notation

\be Z^{\star k}=\underbrace{Z\star \cdots \star Z}_{\mbox{$k$
factors}}\ .\ee
This basis yields the following unique decomposition of $\cP$:

\be \cP=\cP^{(0)}+Z\star \cP^{(1)}+Z^{\star 2}\star
\cP^{(2)}+\cdots\ . \la{ge} \ee
Since $Z$ is central and $\t_\y(Z)=-Z$, it follows from \eq{tc}
that $\t(P^{(k)})=(-1)^{1+k}P^{(k)}$. Hence $\cP^{(k)}$ is
isomorphic to $\cP^{(0)}$ or $\cP^{(1)}$ for $k$ even or odd,
respectively. Since $Z$ is also hermitian, the traceless
multi-spinors $P^{(k)}$ obey the reality condition:

\be \bar{P}^{(k)}{}^{\b_1\dots\b_n\,j_1\dots
j_q}_{\a_1\dots\a_m\,i_1\dots i_p} = -(-1)^{\ft12 p(p-1)+\ft12
q(q-1)} P^{(k)}{}^{\b_1\dots\b_n\,j_1\dots
j_q}_{\a_1\dots\a_m\,i_1\dots i_p}\ ,\la{rc}\ee
where the conjugation is defined as

\be \bar{F}^{\d_1\dots\d_n\,j_1\dots j_q}_{\c_1\dots
\c_m\,i_1\dots
i_p}=(i\C^0)_{\c_1}{}^{\a_1}\cdots(i\C^0)_{\c_m}{}^{\a_m}
\left(F{}^{\a_1\dots\a_m\,i_1\dots i_p}_{\b_1\dots \b_n\,j_1\dots
j_q}\right)^{\dagger} (i\C^0)_{\b_1}{}^{\d_1} \cdots
(i\C^0)_{\b_n}{}^{\d_n} \ .\la{conj}\ee
The degeneracy in $\cP$ due to $Z$ having spin zero suggests that
$Z$ should be eliminated from the higher spin algebra. The Lie
bracket \eq{l} induces a set of brackets with the following
structure:

\be [\cdot,\cdot]\ :\ \cP^{(k_1)}\times \cP^{(k_2)}\ \rightarrow\
\cP^{(k_1+k_2)}+\cP^{(k_1+k_2+1)}+\cdots\ ,\la{l2}\ee
where the direct sum is finite. This structure is due to the fact
that the Lie bracket \eq{l} does not preserve the tracelessness
condition in \eq{alge2}. Thus the higher spin algebra cannot
simply be the restriction of $\cP$ to $\cP^{(0)}$. Instead we let

\be \cI=Z\star \cP^{(1)}+Z^{\star 2}\star \cP^{(2)}+\cdots \ee
This space forms an ideal in $\cP$, i.e. $[\cP,\cI]=\cI$. We next
observe that $U(1)_Y$ acts as an outer automorphism of $\cP$, i.e.
$\cP\backslash U(1)_Y$ forms a Lie superalgebra
\footnote{To see this, we need to show that the generator $Y$
never arises on the right hand side of any two commutators in
$\cP$. Since $Y$ is an $SU(2,2)\times SU(4)$ singlet it can only
arise on the right hand side of graded commutators
$[P_1,\bar{P}_2]$, where $P_1$ is an arbitrary element of $\cP$
and $\bar{P}_2$ its conjugate ($P_1$ and $P_2$ have different
parameters). There is no loss of generality in replacing $P_1$ by
$F=\yb^{\a_1}\cdots \yb^{\a_m}y_{\b_1}\cdots
y_{\b_n}\bar{\th}^{i_1}\cdots \bar{\th}^{i_p}\th_{j_1}\cdots
\th_{j_q}$ (including traces). Computing the part of
$[F,F^\dagger]$ which is quadratic in oscillators, one finds that
the contributions to $K$ and $X$ is given by an overall sign and
combinatorial factor times $(m-n)K+(q-p)X=2(m-n)Z$, where we have
used the $\t_\y$-condition which sets $m-n=q-p$. This shows that
$\cP \backslash U(1)_Y$ is a Lie superalgebra on which $U(1)_Y$
acts as an outer automorphism.}
Thus we can define the higher spin algebra $hs(2,2|4)$ as the
following coset
\footnote{Modding out the ideal generated by
$Z-\ft12 N$, where $N$ is an integer, yields a higher spin
extensions of $SU(2,2|4)$ with central charge $\ft12 N$
\cite{5dv1}.  }:

\be hs(2,2|4)=(\cP\backslash U(1)_Y)/\cI\ .\ee
The elements of $hs(2,2|4)$ are thus equivalence classes $[P]$ of
elements in $\cP$ defined by

\be [P]=\{Q\in \cP\ |\ P-Q\in \cI\}\ .\la{hs}\ee
The Lie bracket of $[P]$ and $[Q]$ is given by

\be [[P],[Q]]=[[P,Q]_*]\ .\ee
In order to exhibit the $SU(4)\times U(1)_Y$ content of the
algebra we define

\be [K,P]=\D P\ ,\quad [X,P]= Y P\ ,\la{dy}\ee
where $\D$ and $Y$, which is the $U(1)_Y$ charge, are the
integers and

\be \D=n_\yb-n_y\ ,\quad Y=n_{\bar{\th}}-n_\th\ ,\la{delta}\ee
with $n_\yb=m$, $n_y=n$, $n_{\bar{\th}}=p$ and $n_\th=q$ defined
by the oscillator expansion given in \eq{alge1}. The condition
\eq{tco} implies that the total $U(1)_Z$ charge vanishes:

\be 2Z=\Delta+Y=0\ .\la{dy2}\ee
The $SU(2)_L\times SU(2)_R$ spins of a generator are given by
$(j_L,j_R)=(\ft12 n_\yb,\ft12 n_y)$. The total spin $s$ of the
gauge field associated with the generator in \eq{alge1} is given
by $s=1+{m+n\over 2}$. From $\t_\y$-condition given in \eq{tc} it
follows that the $SU(4)\times U(1)_Y$ content of the higher spin
algebra is given by:

\bea \ba{l}m+n=0\\s=1\ea&:&\left\{\ba{ll} (0,0)X^3&\qquad Y=0\\
(1,1)(1,X^2)&\qquad Y=0\ea\right.\nn\\[15pt]
\ba{l}m+n=1\\s=\ft32\ea&:& \left\{\ba{ll}
(1,0)(1,X^2)&\qquad Y=1\\(2,1)X&\qquad Y=1\ea\right.\nn\\[15pt]
\ba{l}m+n=2\\s=2\ea&:&\left\{\ba{ll}
(0,0)(1,X^2,X^4)&Y=0\\(1,1)X&Y=0\\(2,2)&Y=0
\\(2,0)X&Y=2\\
(3,1)&Y=2\ea\right.\nn\\[15pt]
\ba{l}m+n=3,7,11,\dots\\s=\ft52,\ft92,\ft{13}2,\dots\ea &:&
\left\{
\ba{ll}(1,0)(X,X^3)&\quad\ Y=1\\(2,1)&\quad\ Y=1\\(3,0)&\quad\ Y=3\ea \right. \nn\\[15pt]
\ba{l}m+n=4,8,12,\dots\\s=3,5,7,\dots\ea&:&\left\{\ba{ll} (0,0)(X,X^3)&\qquad Y=0\\
(1,1)(1,X^2)&\qquad Y=0\\(2,0)(1,X^2)&\qquad Y=2\ea\right.\nn\\[15pt]
\ba{l}m+n=5,9,11\dots\\s=\ft72,\ft{11}2,\ft{15}2,\dots \ea&:&
\left\{\ba{ll}
(1,0)(1,X^2)&\qquad\ Y=1\\(2,1)X&\ \qquad Y=1\\(3,0)X&\ \qquad Y=3\ea\right.\nn\\[15pt]
\ba{l}m+n=6,10,14,\dots\\s=4,6,8,\dots\ea&:&\left\{\ba{ll} (0,0)(1,X^2,X^4)&\quad Y=0\\
(1,1)X&\quad Y=0\\(2,2)&\quad Y=0\\(4,0)&\quad Y=4\ea\right.
\la{algc} \eea
and hermitian conjugates. Here $(p,q)$ denotes the traceless
product of $p$ $\bar{\th}$'s and $q$ $\th$'s. In deriving this
result we have used

\be \left(\bar{\th}^{i_1}\cdots \bar{\th}^{i_p}\th_{i_1}\cdots
\th_{i_q}-{\rm trace}\right)X^{r}=0\quad {\rm for}\quad p+q+r>4\
.\la{pqr}\ee
The content of \eq{algc} is summarized in Tables \ref{td} and
\ref{te}.

\td

\te


\section{The Spectrum of Massless Fields}


We seek an appropriate massless representation $\cS$ of
$hs(2,2|4)$ which will be the spectrum of physical fields in a
field theoretical realization of $hs(2,2|4)$ in five dimensions.
To begin with we observe that each superdoubleton is a
representation of $\cP$. Since all massless representations of the
$N=8$ AdS$_5$ superalgebra consist of the irreps arising in the
tensor products of two superdoubletons, we shall assume that all
massless irreps of $\cP$ also arise in this way. Since we
furthermore have defined the higher spin algebra $hs(2,2|4)$ by
modding out the central charge $Z$, the spectrum $\cS$ must
consist of tensor products between two conjugate spin $j$
superdoubletons with central charges $\ft12 j$ and $-\ft12 j$,
respectively. We next recall the arguments given in
\cite{su1,su2,us}, which suggest that the $hs(2,2|4)$ theory
arises in the near horizon region of weakly coupled, coinciding
D$3$-branes in the large $N$ limit (i.e. in the limit of
tensionless type IIB string theory). If we assume that the higher
spin symmetry is realized as a global conformal symmetry on the
doubletons in the boundary as well, then it is natural to choose
the spectrum $\cS$ to be the symmetric tensor product of two
Yang-Mills superdoubletons (since these are the only doubletons
with $Z=0$), i.e.

\be \cS=(\rm {\cal D}\otimes {\cal D})_S\ ,\la{cs}\ee
where ${\cal D}$ denotes the Yang-Mills superdoubleton:

\be {\cal D}=[D(0,0;1)\otimes 6]_{_0} \oplus
[D(\ft12,0;\ft32)\otimes 4]_{_{-1}} \oplus
[D(0,\ft12;\ft32)\otimes \bar{4}]_{_{1}} \oplus [D(1,0;2)\otimes
1]_{_{-2}} \oplus [D(0,1;2)\otimes 1]_{_2}\ .\ee
The spectrum $\cS$ is listed in Table \ref{tc}. The anti-symmetric
tensor product gives rise to descendants from the boundary CFT
point of view. In fact, the spectrum of the five-dimensional
higher spin gauge theory based on $hs(2,2|4)$ is in one-to-one
correspondence with the algebra of supercurrents of the $d=4$,
$N=4$ Yang-Mills theory. In this case the right hand side of
\eq{cs} is understood to contain a trace over the adjoint $SU(N)$
index, which is a global symmetry of the (free) superdoubleton
theory.


\section{Gauging $hs(2,2|4)$}\la{sec:g}


In order to realize $hs(2,2|4)$ as a local symmetry in a field
theory with spectrum $\cS$ we need to address the issues of
auxiliary gauge fields as well as the incorporation of the
physical spin $s<1$ fields and generalized higher spin
anti-symmetric tensor fields (which describe the $|Y|>1$ sector
of the spectrum exhibited in Table \ref{te}).

Gauging of $hs(2,2|4)$ introduces both dynamic gauge fields and
auxiliary gauge fields. The structure of the set of gauge fields
and curvature constraints, that give rise to one massless spin $s$
degree of freedom on-shell, are known at the linearized level (in
an expansion around $AdS_5$ spacetime). These were first given
using tensors and tensor-spinors \cite{vas1,vas2}. In \cite{us}
this structure was converted into the multi-spinor basis in the
case of $\D=0$ (i.e. bosonic gauge fields). This has an obvious
generalization to $|\D|=1$ (i.e. fermionic gauge fields), which we
shall give below. The gauge fields corresponding to the
generators listed in Table \ref{td} (having $|Y|=|\D|\leq 1$)
therefore give rise to physical spin $s$ degrees of freedom.

The generators listed in Table \ref{te} (having $|\D|=|Y|>1$) do
not give rise to the canonical set of spin $s$ physical and
auxiliary gauge fields. As we shall see these gauge fields
instead have the interpretation of higher spin duals of a certain
tower of higher spin generalizations of the anti-symmetric
two-form potential that arises in the supergravity multiplet.
Thus, comparing the spectrum $\cS$ given in Table \ref{tc} with
the physical states arising from the algebra, as listed in Table
\ref{td}, we find that the spectrum has a `matter content' (the
spin $s<1$ sector and the above mentioned two-form potentials),
given by the states listed in Table \ref{tf}, which cannot be
realized using the gauge field.

\tf

In order to accommodate the spectrum $\cS$ in a field theoretical
construction in five dimensions we introduce a master gauge field
$[A]$ in the adjoint representation of $hs(2,2|4)$, where $A=dx^\m
A_\m$ is a $\cP$-valued one-form, and a master scalar field $\Phi$
in a quasi-adjoint representation $\cR$ of $hs(2,2|4)$ defined as
follows ($|\y|=1$):

\be \t_\y(A)=-A\ ,\quad (A)^{\dagger}=-A\ ,\la{ac}\ee

\be \t_\y(\Phi)=\pi_\y(\Phi)\ ,\quad
(\Phi)^{\dagger}=\pi_{-i}(\Phi)\ , \la{fc}\ee

\be Z\star \Phi = \Phi\star \pi_{-i}(Z)=0\ ,\la{zc}\ee
where the anti-involution $\t_\y$ is defined in \eq{tau} and
$\pi_\y$ is the involution ($|\y|=1$)

\bea
\pi_\y(F(y_\a,\yb^\a,\th_i,\bar{\th}^i))&=&F(-i\bar{\y}\yb_\a,i\y
y^\a,i\y \th_i,-i\bar{\y}\bar{\th}^i)\ ,\\[15pt] \pi_\y(F\star
G)&=&\pi_\y(F)\star \pi_\y(G)\ .\la{i}\eea
The $\cP$ gauge transformations are given by:

\bea
\d_\e A &=& d\e+[A,\e]_*\ ,
\w2
\d_\e \Phi &=&\Phi\star
\pi_{-i}(\e)-\e\star \Phi\ ,
\la{trp}
\eea
where $\e$ is a $\cP$ valued local parameter. The following
curvature and covariant derivative obey \eq{ac} and \eq{fc}

\bea F_A&=&dA+A\wedge\star A\ ,\la{fa}\w2 D_A
\Phi&=&d\Phi-\Phi\star \pi_{-i}(A)+A\star \Phi\ , \la{cd}\eea
and transform in a $\cP$-covariant way as

\bea\d_\e F_A&=&[F_A,\e]_\star\ ,\w2 \d_\e D_A \Phi&=&D_A \Phi
\star\pi_{-i}(\e)-\e\star D_A \Phi\ .  \la{hsgt}\eea

The $\pi_{-i}$-maps are inserted in the definitions of $\d_\e\Phi$
and $D_A\Phi$ to ensure that they obey the $\t$-covariance and
reality conditions given in  \eq{fc}. To show this we first note
that an element $P\in\cP$ obeys

\be \pi_\y^{-1}(P)=\pi_{-i}(P)\ ,\quad
\t_\y(\pi_{-i}(P))=-\pi_\y(P)\ ,\ee
which can be checked explicitly. From this it follows that

\bea &\t_\y\left[\Phi\star\pi_{-i}(P)-P\star\Phi\right]=
\t_\y(\pi_{-i}(P))\star \t_\y(\Phi)-
\t_\y(\Phi)\star \t_\y(P)&\nn\\[15pt]& =-\pi_\y(P)\star
\pi_\y(\Phi)+\pi_\y(\Phi)\star P   = \pi_\y\left[\Phi\star
\pi_{-i}(P)-P\star \Phi\right]&\ ,\nn\eea
hence the $\t$-covariance of $\d_\e\Phi$ and $D_A\Phi$. To check
the reality condition, we first note that $\pi_{-i}^2=1$, which
implies that $(\pi_{-i}(F))^\dagger=\pi_{-i}(F^\dagger)$ for any
function $F$ of the oscillators. It then follows that

\bea& \left[\Phi\star\pi_{-i}(P)-P\star\Phi\right]^\dagger =
(\pi_{-i}(P))^\dagger\star \Phi^\dagger -
\Phi^\dagger \star P^\dagger &\nn\\[15pt]& =-\pi_{-i}(P)\star
\pi_{-i}(\Phi)+\pi_{-i}(\Phi)\star P   = \pi_{-i}\left[\Phi\star
\pi_{-i}(P)-P\star \Phi\right]&\ ,\nn\eea
hence the reality conditions of $\d_\e\Phi$ and $D_A\Phi$. The
$hs(2,2|4)$-valued gauge field, curvature, covariant derivative
and gauge parameter are defined by

\be [A]\ ,\quad F_{[A]}=[F_A]\ ,\quad D_{[A]}\Phi=D_A\Phi\ ,\quad
[\e]\ ,\ee
where we use the notation defined in \eq{hs}. The $hs(2,2|4)$
gauge transformations are

\be \d_{[\e]}[A]=[\d_\e A]\ ,\quad \d_{[\e]}F_{[A]}=[\delta_\e
F_A]\ ,\quad \d_{[\e]}\Phi=\d_\e\Phi\ .\ee
The condition \eq{zc} implies that $D_{A}\Phi$ and $\d_{\e}\Phi$
are independent of the choice of $\cP$-valued representatives $A$
and $\e$. The curvature $F_{[A]}$ and the gauge transformation
$\d_{[\e]}[A]$ are computed by first evaluating the ordinary
$\star$ product between the representatives and then expanding
the result with respect to the particular ordering of oscillators
defined by \eq{ge} and finally discarding any terms in $\cI$. In
case one would have to perform several repeated multiplications
of objects in $hs(2,2|4)$ the last step may of course be carried
out at the end, as the operation of modding out $Z$ commutes with
taking the $\star$ product.

As found in the previous section, the gauge field $[A]$ can be
represented by $A\in \cP^{(0)}$, which has an expansion in terms
of component fields $A_{\m\,}{}^{\b_1\dots \b_n}_{\a_1\dots
\a_m}{}^{j_1\dots j_q}_{i_1\dots i_p}$, with tangent indices
corresponding to the generators given in \eq{alge1} and
\eq{alge2}. The $SU(4)$ content is listed explicitly in
\eq{algc}; see also Tables \ref{td} and \ref{te}.


\section{The Field Content of $\Phi$}


The field content of the master field $\Phi$ is determined by the
$\t_\y$ and reality conditions in \eq{fc} and the $Z$-condition in
\eq{zc}. To solve these conditions it is convenient to use the
notation:

\be \Phi^{(r;t)}(m,n;p,q) = {1\over
m!n!p!q!}\Phi^{(r;t)}{}^{\b_1\dots \b_n}_{\a_1\dots
\a_m}{}^{j_1\dots j_q}_{i_1\dots i_p} \yb^{\a_1}\cdots
\yb^{\a_m}y_{\b_1}\cdots y_{\b_n}\bar{\th}^{i_1}\cdots
\bar{\th}^{i_p}\th_{j_1}\cdots \th_{j_q} \ ,\la{rt}\ee
where the superscript $r$ denotes the number of anti-symmetric
pairs of spinor indices and $t$ ($0\leq t\leq r$) the number of
these that are traced. The spin of a component field of $\Phi$ is
defined by $s={m+n\over 2}-t$. The relation between the labels $s$
and $r$ and the $SO(5)$ highest weights $(m_1,m_2)$ ($m_1\geq
m_2\geq 0$) of the irreducible component $T^{(m_1,m_2)}$
contained in $\Phi^{(r;t)}$ are given in \eq{hwrs}. Below we
shall most frequently use the following components:

\bea \Phi^{(0;0)}_{\a_1\dots\a_m,\b_1\dots\b_n}&=&
T^{(\fft{m+n}2,\fft{m+n}2)}_{\a_1\dots\a_m\b_1\dots\b_n}\ ,
\nn\w2
\Phi^{(1;0)}_{\a_1\dots\a_m,\b_1\dots\b_n}&=&
T^{(\fft{m+n}2,\fft{m+n-2}2)}_{\a_1\dots\a_m\b_1\dots\b_n}\ ,
\nn\w2
\Phi^{(1;1)}_{\a_1\dots\a_m,\b_1\dots\b_n}&=&
C^{\phantom{\fft{m}2}}_{\a_1\b_1}
T^{(\fft{m+n-2}2,\fft{m+n-4}2)}_{\a_2\dots\a_m\b_1\dots\b_n}\ ,
\nn\w2
\Phi^{(2;0)}_{\a_1\dots\a_m,\b_1\dots\b_n}&=&
T^{(\fft{m+n}2,\fft{m+n-4}2)}_{\a_1\dots\a_m\b_1\dots\b_n}\ ,
\nn\w2
\Phi^{(2;1)}_{\a_1\dots\a_m,\b_1\dots\b_n}&=&
C^{\phantom{\fft{m}2}}_{\a_1\b_1}
T^{(\fft{m+n-2}2,\fft{m+n-4}2)}_{\a_2\dots\a_m\b_1\dots\b_n}\ ,
\nn\w2
\Phi^{(2;2)}_{\a_1\dots\a_m,\b_1\dots\b_n}&=&
C^{\phantom{\fft{m}2}}_{\a_1\b_1}
C^{\phantom{\fft{m}2}}_{\a_2\b_2}
T^{(\fft{m+n-4}2,\fft{m+n-4}2)}_{\a_3\dots\a_m\b_1\dots\b_n} \
,\nn\eea
where the separate symmetrizations in $\a$ and $\b$-indices on
the right hand sides have been suppressed. The $\t_\y$-condition
implies

\be m+n+p+q=\left\{\ba{lc}0\mbox{ mod $2$}& \mbox{ for }m\neq
n\\2r\mbox{ mod $4$}& \mbox{ for }m=n\ea\right.\la{mnpq}\ee
 From this it follows that if $\Phi^{(0;0)}(m,n;p,q)$ is allowed,
then $\Phi^{(r;t)}(m+r,n+r;p,q)$n with $r,t=0,1,\dots$ is also
allowed. It is therefore sufficient to analyze the $SU(4)$ content
for $r=t=0$. Using \eq{pqr}, we find that for $m=n$ the possible
$SU(4)$ contents are ($s=\ft{m+n}2$):

\bea \ba{l} m=n=0,2,\dots\\ s=0,2,\dots \ea &:&\left\{\ba{l} (0,0)(1,X^2,X^4)\\
(2,2)\\ (1,1)X\\ (3,1)\\ (2,0)X\\(4,0)\ea\right.\nn\\
\ba{l} m=n=1,3,\dots\\ s=1,3,\dots \ea &:&\left\{\ba{l}
(0,0)(X,X^3)\\(1,1)(1,X^2)\\(2,0)(1,X^2)\ea\right.\la{men}\eea
and hermitian conjugates, where $(p,q)$ denotes the traceless
product of $p$ number of $\bar{\th}$'s and $q$ number of $\th$'s.
For $m\neq n$ we find

\bea \ba{l} m+n=1,3,\dots\\s=\ft12,\ft32,\dots\ea
&:&\left\{\ba{l}  (1,0)(1,X,X^2,X^3)\\
(2,1)(1,X)\\(3,0)(1,X)\ea\right.\nn\\
\ba{l} m+n=2,4,\dots\\s=1,2,\dots \ea
&:&\left\{\ba{l}  (0,0)(1,X,X^2,X^3,X^4)\\ (1,1)(1,X,X^2)\\
(2,0)(1,X,X^2)\\(4,0)\\(3,1)\\(2,2) \ea\right.\la{mnen}\eea
and hermitian conjugates. The reality condition in \eq{fc}
determines $\Phi(m,n;...)$ in terms of $\Phi(n,m;...)$.

We next examine the $Z$-condition \eq{zc}. Adding and
subtracting the two equations, we obtain

\bea \{X,\Phi(m,n;p,q)\}_\star +\D \Phi(m,n;p,q)&=&0\
,\la{xe}\\[15pt]
\{K,\Phi(m,n;p,q)\}_\star +Y \Phi(m,n;p,q)&=&0\ ,\la{ke}\eea
where $[K,\Phi]_\star=\D\Phi$ and $[X,\Phi]_\star=Y\Phi$
\footnote{The $U(1)_Y$-charge of $\Phi$ is
given by
$Y\Phi:=X\star\Phi-\Phi\star\pi_{-i}(X)=[X,\Phi]_\star$.}.
The anti-commutators in \eq{xe} and \eq{ke} are evaluated using

\bea K\star \left(K^t X^s
T(m,n;p,q)\right)&=&\left(K^{t+1}+\ft12\D
K^t-\ft14 t(t+3+m+n)K^{t-1}\right)X^sT(m,n;p,q)\ ,\nn\\[15pt]
\left(K^t X^sT(m,n;p,q)\right)\star K &=&\left(K^{t+1}-\ft12\D
K^t-\ft14
t(t+3+m+n)K^{t-1}\right)X^sT(m,n;p,q)\ ,\nn\\[15pt]
X\star \left(K^t X^s T(m,n;p,q)\right)&=&\left(X^{s+1}-\ft12Y
X^s-\ft14 s(s-5+p+q)X^{s-1}\right)K^tT(m,n;p,q)\ ,\nn\\[15pt]
\left(K^t X^s T(m,n;p,q)\right)\star X &=&\left(X^{s+1}+\ft12Y
X^s-\ft14 s(s-5+p+q)X^{s-1}\right)K^tT(m,n;p,q)\ ,\nn \eea
where $T(m,n;p,q)$ is assumed to be traceless both in $SU(2,2)$
and $SU(4)$ indices. Eq. \eq{xe} gives a characteristic equation
for $\D$ and furthermore eliminates some of the $SU(4)$
representations listed in \eq{men} and \eq{mnen}. We find
($Y=p-q$ for $(p,q)X^r$ elements; $\D=m-n$):

\bea
\ba{l}m=n=0,2,\dots\\s=0,2,\dots\ea&:&\left\{\ba{l} (0,0)(1-\ft23
X^2+\ft23X^4)\\(2,2)\\ (3,1)\\(4,0)
\ea\right.\nn\\[15pt]
\ba{l}m=n=1,3,\dots\\s=1,3,\dots\ea&:&\left\{\ba{l} (1,1)(1-\ft12
X^2)\\(2,0)(1-\ft12 X^2)\ea\right.\nn\\[15pt]
\ba{l}m+n=1,3,\dots\\s=\ft12,\ft32,\dots\ea &:& \left\{ \ba{ll}
(1,0)\left(\ba{c}1-\ft{2\D}{3} X +\ft13(\D^2-3)X^2\\
-\ft{2}{3\D}(\D^2-3)X^3\ea\right)
\qquad&\D=\pm1,\pm3\\[15pt]
(2,1)(1-2\D X)&\D=\pm1\\
(3,0)(1-2\D X)&\D=\pm1\ea \right. \nn\\[15pt]
\ba{l} m+n=2,4,\dots\\s=1,2,\dots\ea  &:& \left\{\ba{ll}
(0,0)\left(\ba{c}1-\ft{\D}{2}X+\ft16(\D^2-4)X^2\\-\ft{\D}{18}
(\D^2-10)(1-\ft{2}{\D}X)X^3\ea\right)\quad&
\D=\pm2,\pm4\\[15pt]
(1,1)(1-\D X+2X^2)&\D=\pm2\\
(2,0)(1-\D X+2X^2)&\D=\pm2 \ea\right.\la{phicontent}\eea
and hermitian conjugates (which have opposite values of $Y$). The
condition \eq{ke} relates components of $\Phi$ that have different
powers of $K$. Denote such structures by $\Phi_t K^t
=\Phi^{(r+t;t)}(m,n;p,q)$ and expand

\be \Phi(K)=\sum_{t=0}^\infty \Phi_t K^t\ .\la{kexp1}\ee
 From \eq{ke} it follows that

\be 2\Phi_{t-1}+Y \Phi_t -\ft{t+1}2(t+4+p+q)\Phi_{t+1}=0\
.\la{kexp2}\ee
This equation determines $\Phi_t$ uniquely in terms of the
leading coefficient $\Phi_0$:

\be \Phi(K)=f(m+n,Y;K)\Phi_0\ ,\la{kexp3}\ee
where the function $f(m+n,Y;z)$ is an analytic function. In the
analysis of the linearized field equations one only needs to
expand up to first order in $K$, except in the scalar sector
where also the second order is needed. The first order
coefficients are given by

\be \Phi^{(r+1;1)}_{\a_1\dots\a_m,\b_1\dots\b_n}(\th,\bar{\th})=
Y{mn\over m+n+2}
C^{\phantom{^{(r;0)}}}_{\a_1\b_1}\Phi^{(r;0)}_{\a_2\dots\a_{m},\b_2\dots
\b_{n}}(\th,\bar{\th})\
,\la{tp}\ee
where separate symmetrization of the $\a$ and $\b$ indices is
understood. In the case of scalars the $K$-expansion reads:

\be \phi(K)=\left(1+\ft{Y}{
2}K+(\ft25+\ft{Y^2}{10})K^2+\cdots\right)\phi_0\ .\la{phk}\ee

In summary, the field content of $\Phi$ is arranged into $K$ and
$X$ expansions starting with $SU(2,2)$ traceless multi-spinors
$\Phi^{(r;0)}(m,n;p,q)$ whose $SU(4)$ content is given by
\eq{phicontent}. As we shall see in the next section, the
expansion of the anti-symmetric traceless $SU(2,2)$ indices
yields trajectories of the schematic form
$$\Phi^{(r;0)}(m,n;p,q)\sim P^{a_1}\cdots
P^{a_r}\nabla_{a_1}\cdots \nabla_{a_r}\Phi^{(0;0)}(m-r,n-r;p,q)\ ,
$$ so that only $\Phi^{(0;0)}(m,n;p,q)$ are independent fields.
The sector with $s\geq 1$ and $|Y|\leq 1$ are Weyl tensors while
the sector with $s\geq 1$ and $|Y|>1$ are independent three-form
field strengths (carrying internal tangent space indices when
$s>1$) obeying self-duality equations in $D=5$. The $s<1$ sector
defines physical scalars and fermions. Thus the independent field
content of $\Phi$ matches that of the spectrum in Table \ref{tc}
(though it remains to verify that the lowest energies satisfy
\eq{e0}). By construction, the components of $\Phi$ fall into
supermultiplets labeled by the level index of \eq{alge1}:

\bea \Phi&=&\sum_{\ell=0,1,2,\dots} \Phi_{(\ell)}\la{ell}\ ,\\[15pt]
\d_\e\Phi_{(\ell)}&=&\e\star\Phi_{(\ell)}-\Phi_{(\ell)}\star\pi_{-i}(
\e)\ ,\la{susy}\eea
where $\e=\e_\a^i\yb^\a\th_i-h.c$ denotes the supersymmetry
parameter. The level index breaks the degeneracy whenever there
are several components of $\Phi$ carrying the same $SU(2,2)\times
SU(2)\times U(1)_Y$ representation.

The spin $s\leq 1$ sector is embedded in $\Phi$ as follows:

\bea \Phi&=& \phi(\th,\bar{\th};K) + y^\a \l_\a(\th,\bar{\th};K) +
\yb^\a \tilde{\l}_\a(\th,\bar{\th};K) \nn\\[15pt] && +\ft12 y^\a y^\b
\phi_{\a\b}(\th,\bar{\th};K)+ \ft12 \yb^\a\yb^\b
\tilde{\phi}_{\a\b}(\th,\bar{\th};K)\nn\\[15pt] && +
y^{(\a}\yb^{\b)}G_{\a\b}(\th,\bar{\th};K)+ i\yb \C^a y
\phi_{a}(\th,\bar{\th};K) +\cdots\ ,\la{ex}\eea
where

\bea \phi(\th,\bar{\th};K)&=&\ft14
\bar{\th}^i\bar{\th}^j\th_k\th_l
\phi_{ij}^{kl}(K)+\ft16\bar{\th}^i\bar{\th}^j\bar{\th}^k\th_l
\phi_{ijk}^l(K)+\ft1{24}\bar{\th}^i\bar{\th}^j\bar{\th}^k\bar{\th}^l
\phi_{ijkl}(K)\nn\\[15pt]&&
+ (1-\ft23 X^2+\ft23X^4)\phi(K) + \mbox{conj.}\ , \nn\\[15pt]
\l_\a(\th,\bar{\th};K)&=&(1+2X)\left[\ft12
\bar{\th}^i\bar{\th}^j\th_k \l_{\a}{}_{ij}^k(K) + \ft16
\bar{\th}^i\bar{\th}^j\bar{\th}^k\l_\a{}_{ijk}(K)\right] \la{slo}\\[15pt] &&+
(1+\ft23X-\ft23X^2-\ft43 X^3)\bar{\th}^i\l_{\a\,i}(K)+ \mbox{conj.}\ ,\nn\\[15pt]
\phi_{\a\b}(\th,\bar{\th};K)&=& \ft12 (\C^{ab})_{\a\b}\left[
(1+2X+2X^2)\bar{\th}^i\th_j F_{ab}{}_i^j(K)
+(1+2X-\ft23X^3-\ft23X^4)F_{ab}(K)\right]\nn\\[15pt]&& +\ft1{12}
(\C^{abc})_{\a\b}  (1+2X+2X^2)\bar{\th}^i\bar{\th}^j
H_{abc\,ij}(K)+ \mbox{conj.}
\ ,\nn\\[15pt]
G_{\a\b}(\th,\bar{\th};K)&=& (1-\ft12X^2)\left[\ft12
(\C^{ab})_{\a\b} \bar{\th}^i\th_j F'_{ab}{}_i^j(K) + \ft1{12}
(\C^{abc})_{\a\b}  \bar{\th}^i\bar{\th}^j H'_{abc\,ij}(K)\right]+
\mbox{conj.}\ .\nn \eea
Here all $SU(4)$ representations are irreducible and the
$K$-dependence, which is of the form \eq{kexp3}, is determined by
\eq{kexp2}. The conjugates are determined by the reality
conditions:

\bea (\phi(\th,\bar{\th}))^{\dagger}&=&\phi(\th,\bar{\th})\ ,\quad
(\l_\a(\th,\bar{\th}))^{\dagger}=-\bar{\l}_\a(\th,\bar{\th})\
,\nn\\[15pt]
(\phi_{\a\b}(\th,\bar{\th}))^{\dagger}&=&
\bar{\phi}_{\a\b}(\th,\bar{\th})^{\dagger}\
,\quad (G_{\a\b}(\th,\bar{\th}))^\dagger =
\bar{G}_{\a\b}(\th,\bar{\th})\ ,\eea
where $\bar{\l}_\a$, $\bar{\phi}_{\a\b}$ and $\bar{G}_{\a\b}$ are
defined by \eq{conj}. The $\t_\y$-condition implies
$\tilde{\l}_\a(\th,\bar{\th})=i\l_\a(i\th,i\bar{\th})$ and
$\tilde{\phi}_{\a\b}(\th,\bar{\th})=-\phi_{\a\b}(i\th,i\bar{\th})$.
The $SU(4)$ content and reality condition of
$\phi_a(\th,\bar{\th})$ is the same as that of
$\phi(\th,\bar{\th})$. The linearized field equation will set
$\phi_a(\th,\bar{\th})\sim \partial_a \phi (\th,\bar{\th})$ and
equate $F_{ab}{}_i^j$, $F'_{ab}{}_i^j$ and $F_{ab}$ with the
curvatures of the corresponding spin $1$ gauge fields, while
$H_{abc\,ij}$ and $H'_{abc\,ij}$ will be field strengths of
physical two-form potentials $B_{\m\n\,ij}$ and $B'_{\m\n\,ij}$.
The level $\ell=1$ fields are $\phi$, $\l_{\a i}$, $F_{ab}$,
linear combinations of $F_{ab}{}_i^j$ and $F'_{ab}{}_i^j$, and
linear combinations of $H_{abc\,ij}$ and $H'_{abc\,ij}$,
determined by \eq{susy}.


\section{The Linearized Constraints and Their Integrability}


We assume that the higher spin gauge theory can be expanded
around the AdS-vacuum described $\Phi=0$ and $A=\Omega$. Building
upon the bosonic results in \cite{us}, we propose the following
linearized constraints:

\bea F_{(\ell)\a_1\dots\a_m,\b_1\dots\b_n}{}_{i_1\dots
i_p}^{j_1\dots j_q}&=&e^a\wedge
e^b(\C_{ab})^{\c\d}\Phi^{(0;0)}_{(\ell)\c\a_1\dots\a_m,\d\b_1\dots
\b_n}{}_{i_1\dots
i_p}^{j_1\dots j_q}\mid_{\cP^{(0)}\backslash U(1)_Y}\ ,\la{feph}\\[15pt]
d\Phi+\Omega\star\Phi-\Phi\star\pi_{-i}(\Omega)&=&0\
,\la{dphez}\eea
where $\Phi^{(0;0)}$ is the fully symmetric part of $\Phi$, as
defined in \eq{rt}, and $F$ is the $AdS_5$ covariant linearized
field strength of the $\cP^{(0)}$-valued gauge field $A$:

\bea F&=&dA + \O\star A-A\star \O\
,\nn\\[15pt]
F_{\m\n\,\a_1\dots\a_m,\b_1\dots\b_n}&=&
2\nabla_{[\m}A_{\n]\,\a_1\dots\a_m,\b_1\dots\b_n}\\&&
+m(\C_{[\m})_{\a_1}{}^{\c}A_{\n]\,\c\a_2\dots\a_m,\b_1\dots\b_n}-
n(\C_{[\m})_{\b_1}{}^{\c}A_{\n]\,\a_1\dots\a_m,\c\b_2\dots\b_n} \
.\la{fcomp}\nn\eea
Here $\nabla_\m$ is the background Lorentz covariant derivative,
$SU(4)$ indices are suppressed and separate symmetrizations in
$\a$ and $\b$ indices are understood. In \eq{feph} the subscript
$\ell=0,1,2,...$ refers to the supermultiplet level index defined
in \eq{alge1} and \eq{ell}, and the $SU(4)$ indices are assumed to
be irreducible (which means that $m$, $n$, $p$ and $q$ do not in
general obey the condition in \eq{alge1}). Below, in verifying
integrability and finding the linearized field equations, we can
suppress both the $SU(4)$ indices and the level index, due to the
linear nature of all equations. For $m=n$ and suitable
restrictions on the $SU(4)$ content (see \cite{us}), the equations
\eq{feph} and \eq{dphez} are precisely the bosonic equations
proposed in \cite{us}.

The right hand side of \eq{feph} is a projection of
$\Phi^{(0;0)}$, both in its $(y,\yb)$ and
$(\th,\bar{\th})$-expansion. The components projected out are the
$s<1$ sector and the underlined $SU(4)$ irreps shown in Table
\ref{tf}. While this, as well as the matching of $K$ and
$X$-dependence, is done here by hand, we expect that these
operations arise naturally in the full interacting theory, as is
the case in four dimensions, such that \eq{feph} can be written
as a master constraint $F={\cal V}(\Phi)$, where ${\cal V}$ is a
map involving the background f\"unfbein and the generator $X$.
The linearized constraint \eq{dphez}, on the other hand, is
already in the desired form.

Before we analyze the consequences of \eq{feph} and \eq{dphez} for
the linearized field equations, we first wish to establish their
integrability. Using $d\O+\O\star\O=0$, it immediately follows
that \eq{dphez} is integrable. The integrability of \eq{feph}
requires the right hand side to obey the Bianchi
identity\footnote{Here and in the remainder of this section the
separate symmetrizations in $\a$ and $\b$ indices is understood.}

\bea &dF+\O\star F-F\star\O=0&\ ,\la{bi}\\[15pt]
&\nabla_{[\m}F_{\n\r]\,\a_1\dots\a_m,\b_1\dots\b_n}+ {m\over
2}(\C_{[\m})_{\a_1}{}^\c F_{\n\r]\,\c
\a_2\dots\a_m,\b_1\dots\b_n}- {n\over 2}(\C_{[\m})_{\b_1}{}^\c
F_{\n\r]\, \a_1\dots\a_m,\c\b_2\dots\b_n}=0&\ ,\la{bic}\eea
For $m=n$ this was shown in \cite{us}. To show this in general we
first write the constraints \eq{feph} and \eq{dphez} in
components as follows:

\bea F_{\m\n\,\a_1\dots\a_m,\b_1\dots\b_n}&=& {1\over 8}
(\C_{\m\n})^{\c\d}\Phi^{(0;0)}_{\c\a_1\dots\a_m,\d\b_1\dots\b_n}
\ ,\la{feph1}\\[15pt]
\nabla_\m\Phi_{\a_1\dots\a_m,\b_1\dots\b_n}&=&{1\over 2}
(\C_\m)^{\c\d} \Phi_{\c\a_1\dots\a_m,\d\b_1\dots\b_n}-{mn\over 2}
(\C_\m)_{\a_1\b_1} \Phi_{\a_2\dots\a_m,\b_2\dots\b_n}\ .
\la{dphez1}\eea
We next substitute \eq{feph1} into \eq{bic} and use \eq{dphez1}.
Upon decomposing the free $\a$ and $\b$ indices into $SU(2,2)$
irreps, all structures vanish identically except the $(0;0)$ and
$(1;0)$ parts, where we use the notation of \eq{rt}. The $(1;0)$
part vanishes due to the following Fierz identity \cite{us}:

\be
(\C_{a[\m})^{\b\c}(\C_{\n\r]})^{\d\e}\Phi^{(0;0)}_{\b\c\d\e\a_1\dots
\a_{m+n-2}}=0\
.\la{fi1}\ee
The $(0;0)$ part of \eq{bic} takes the form

\be {1\over
2}(\C_{[\m\n}^{\b\c}(\C_{\r]})^{\d\e}\left[\Phi^{(1;0)}_{\b\c\d\e\a_1\dots
\a_{m+n}}+
\Phi^{(1;1)}_{\b\c\d\e\a_1\dots\a_{m+n}}\right] + {m-n\over
2}(\C_{[\m})_{\a_1}{}^\b(\C_{\n\r]})^{\c\d}\Phi^{(0;0)}_{\b\c\d\a_2\dots
\a_{m+n}}=0\
,\la{bi00}\ee
where $\Phi^{(1;1)}$ is given in terms of $\Phi^{(0;0)}$ by
\eq{tp}, with the $U(1)_Y$ charge $Y=n-m$ (the charge of
$\Phi^{(0;0)}$ is the same as that of the field strength, which in
turn is given by $Y=-\D=n-m$, as follows from \eq{dy2}). The
contribution to \eq{bi00} from $\Phi^{(1;0)}$ vanishes by means
of the following Fierz identity \cite{us}:

\be
(\C_{[\m\n}^{\b\c}(\C_{\r]})^{\d\e}\Phi^{(1;0)}_{\b\c\d\e\a_1\dots\a_{m+n}}=0\
.\la{fi2}\ee
Thus, after some algebra
\footnote{In order to substitute \eq{tp}
into \eq{bi00} one needs to first split the indices
$\a_1\dots\a_{m+1}$ and $\b_1\dots\b_{n+1}$ into
$\a_1\dots\a_m\c\e$ and $\a_1\dots\a_n\d\vf$ and then contract
with $(\C_{[\m})^{\c\d}(\C_{\n\r]})^{\e\vf}$ and finally
symmetrize the remaining free indices.}
, one finds that \eq{bi00} is equivalent to

\bea& Y\left\{{1\over m+n+6}\left((\C_{\m\n\r]})^{\b\c}
\Phi^{(0;0)}_{\b\c\a_1\dots\a_{m+n}}+{m+n\over 2}(\C_{[\m})_{\a_1}{}^{\b}
(\C_{\n\r]})^{\c\d}
\Phi^{(0;0)}_{\b\c\d\a_2\dots\a_{m+n}}\right)\right.&\nn\\[15pt]
&\left.-\ft12 (\C_{[\m})_{\a_1}{}^{\b} (\C_{\n\r]})^{\c\d}
\Phi^{(0;0)}_{\b\c\d\a_2\dots\a_{m+n}}\right\}=0\ ,&\la{last}\eea
where the first two terms come from $\Phi^{(1;1)}$. This equation
simplifies as
\footnote{This condition does not arise in the
bosonic case \cite{us}, since $Y=0$ in that case.}:

\be Y\left\{(\C_{\m\n\r})^{\b\c}
\Phi^{(0;0)}_{\b\c\a_1\dots\a_{m+n}}-3(\C_{[\m})_{(\a_1}{}^{\b}
(\C_{\n\r]})^{\c\d}
\Phi^{(0;0)}_{\a_2\dots\a_{m+n})\b\c\d}\right\}=0\ .\la{last1}\ee
This is satisfied due to the following Fierz identity:

\be \left[(\C_{ab})^{\b\c}(\C_{\m\n\r})^{\d\e}-
3(\C_{[\m\n})^{\b\c}(\C_{\r]ab})^{\d\e}\right]
\Phi^{(0;0)}_{\b\c\d\e\a_1\dots\a_{m+n-2}}=0\ ,\la{fi3}\ee
which follows from the five-dimensional membrane identity:

\be (\C^{\m\n})_{(\a\b}(\C_\n)_{\c)\d}=0\ .\la{fi4}\ee
To apply \eq{fi3} to \eq{last1}, we contract pairs of $\a$-indices
by second rank $\C$-matrices, and symmetrize all pairs of
anti-symmetric vector indices. In the case of odd $m+n$, one in
addition has to make use of the Fierz identity
$$(\C^{[\m})_{[\e}{}^\b(\C^{\n\r]})^{\c\d}\Phi^{(0;0)}_{\varphi]\a_1
\dots\a_{m+n-2}\b\c\d}=0\ ,$$ which follows from \eq{fi1}.

Thus we have established the integrability of the constraints
\eq{feph} and \eq{dphez} describing the linearization of the
$hs(2,2|4)$ gauge theory around $AdS_5$ spacetime.


\section{The Linearized Field Equations}



\subsection{The Master Scalar Constraint}


We begin by analyzing the constraint \eq{dphez} on the master
scalar field, since we expect all physical degrees of freedom of
the theory to be represented in $\Phi$. The physical spin $s\geq
1$ fields occur in $\Phi$ via their field strengths or their
derivatives. Interestingly, as we shall see below, a subset of
these are three-form field strengths obeying self-duality
equations in five dimensions. These are higher spin
generalizations of the well-known field equation for the two-form
potential in the supergravity multiplet, and yield states in the
spectrum with $|Y|>1$ (the remaining spin $s\geq 1$ states, which
has $|Y|\leq 1$, occur in $\Phi$ via their Weyl tensors).

The master scalar constraint \eq{dphez1} implies that
$\Phi^{(r;0)}_{\a_1\dots\a_{2s+2r}}$ ($r\geq 1$) can be expressed
in terms of (up to $r$) derivatives of
$\Phi^{(0;0)}_{\a_1\dots\a_{2s}}$:

\be (\C_{a_1})^{\a_1\a_2}\cdots
(\C_{a_r})^{\a_{2r-1}\a_{2r}}\Phi^{(r,0)}_{\a_1\dots\a_{2r}\b_1\dots\b_{2s}}=2^r\nabla_{(a_1}\cdots\nabla_{a_r)}
\Phi^{(0;0)}_{\b_1\dots\b_{2s}}-\mbox{traces}\ .\ee
In what follows we shall therefore focus on obtaining the field
equations obeyed by the components of $\Phi^{(0;0)}$. Clearly,
since $\Phi$ is a representation space of $hs(2,2|4)$, as given
by \eq{hsgt}, and the field content of $\Phi$ is in one-to-one
correspondence with the massless spectrum listed in Table
\ref{tc}, the lowest energies $E_0$ of the components in
$\Phi^{(0;0)}$ must be given by \eq{e0}, as we shall verify
explicitly using the field equations derived below. In deriving
lowest energy labels $E_0$ from the various physical equations we
use the harmonic analysis described in Appendix B.


\subsubsection*{The Matter Field Equations ($s=0,\ft12$)}


The scalar Klein-Gordon equations follow from the following
components of the scalar master equation \eq{dphez1} \cite{us}

\bea \partial_\m\Phi^{(0;0)} &=& \ft12 (\C_\m)^{\a\b}
\Phi^{(1;0)}_{\a,\b}\ ,\la{se3}\\[15pt]
\nabla_\m\Phi_{[\a,\b]} &=& \ft12 (\C_\m)^{\c\d}\left[
\Phi^{(2;0)}_{[\a|\c,|\b]\d}+\Phi^{(2;1)}_{[\a|\c,|\b]\d}+
\Phi^{(2;2)}_{[\a|\c,|\b]\d}\right]- \ft12
(\C_\m)_{\a\b}\Phi^{(0;0)}\ ,\la{se4}\eea
where all components are coefficients in the expansion \eq{rt}.
The second equation decomposes as

\bea \nabla_\m\Phi^{(1;0)}_{\a,\b} &=& \ft12 (\C_\m)_{\c\d}\left[
\Phi^{(2;0)}_{[\a|\c,|\b]\d} +\Phi^{(2;2)}_{[\a|\c,|\b]\d}\right]-
\ft12
(\C_\m)_{\a\b}\Phi^{(0;0)}\ ,\la{se1}\\[15pt]
\nabla_\m\Phi^{(1;1)}_{\a,\b} &=& \ft12 (\C_\m)_{\c\d}
\Phi^{(2;1)}_{[\a|\c,|\b]\d}\ .\la{se2}\eea
 From the expansion \eq{phk} we read off

\bea \Phi^{(2;1)}_{\a\c,\b\d}&=& \ft{2Y}3
\Phi^{(1;0)}_{(\a|(\b|}C_{|\c)|\d)}\ ,\la{phe1}\\[15pt]
\Phi^{(2;2)}_{\a\c,\b\d} &=& \left(\ft25 + \ft{Y^2}{10}\right)
C_{(\a|(\b|}C_{|\c)|\d)}\Phi^{(0;0)}\ .\la{phe2}\eea
Taking the divergence of \eq{se3} and using \eq{se1} and
\eq{phe2} and the Fierz identity

\be (\C^a)^{\a\b}(\C_a)^{\c\d}\F^{(2,0)}_{\a\b\c\d}=0\ ,\ee
we find

\be (\nabla^\m\partial_\m + 4-\ft{Y^2}4)\Phi^{(0;0)}=0\ .\ee
 From this it follows that the lowest energy is $E_0=2+\ft12|Y|$,
in accordance with \eq{e0}. Thus, at level $\ell=0$, the
$20'_0$-plet, $10_2$-plet and complex $1_4$-plet have $E_0=2$,
$3$ and $4$, respectively, and at level $\ell=1$ the real
$1_0$-plet has $E_0=2$. The remaining equation \eq{se2} is an
identity, upon the use of \eq{phe1}, and $\F^{(2,0)}_{\a\b\c\d}$
contains the non-vanishing second derivatives.

The spin $s=\ft12$ field equations follow from the following
components of the scalar master equation \eq{dphez1}:

\be \nabla_\m \Phi_\a =\ft12 (\C_\m)^{\b\c}\Phi_{\a\b,\c} = \ft12
(\C_\m)^{\b\c}\left[\Phi^{(1;0)}_{\a\b,\c}+\ft{2Y}5
\Phi_{(\a}C_{\b)\c}\right]\ ,\ee
where we have used \eq{tp}. The $\C^\m$-trace yields the Dirac
equation

\be
\left((\C^\m)_\a{}^\b\nabla_\m-\d_\a^\b\ft{Y}2\right)\Phi_\b=0\
.\ee
This gives the lowest energy $E_0=2+\ft12|Y|$, in accordance with
\eq{e0}. Thus, at level $\ell=0$ the $20_1$-plet and the
$4_3$-plet have $E_0=\ft52$ and $\ft72$, respectively, and at
level $\ell=1$ the $4_1$-plet has $E_0=\ft52$. The non-vanishing
derivatives are contained in
$\Phi^{(1;0)}_{\a\b,\c}=(\C^a)_{\c(\a}\Psi_{a\,\b)}$, where
$\Psi_{a\,\a}$ is a $\C$-traceless vector-spinor.


\subsubsection*{The Self-Duality Equations ($s\geq 1$)}


The part of \eq{dphez1} that is totally symmetric in $\a$ and
$\b$ indices can be written as

\be \nabla_\m \Phi^{(0;0)}_{\n\r\,\a_1\dots\a_{2s-2}}=
\ft12(\C_{\n\r})^{\b\c}(\C_\m)^{\d\e}\left[\Phi^{(1;0)}_{\b\c\d\e\a_1\dots
\a_{2s-2}}+
\Phi^{(1;1)}_{\b\c\d\e\a_1\dots\a_{2s-2}}\right]\ ,\la{sd1}\ee
where we use the notation of \eq{rt} and we have defined

\be \Phi^{(0;0)}_{\m\n\,\a_1\dots\a_{2s-2}}=
(\C_{\m\n})^{\b\c}\Phi^{(0;0)}_{\b\c\a_1\dots\a_{2s-2}}\
.\la{sd2}\ee
 From this definition and the membrane identity \eq{fi4}, it
follows that

\be (\C^\m)_\b{}^\c\Phi^{(0;0)}_{\m\n\,\c\a_1\dots\a_{2s-2}}=0\
.\la{sd3}\ee
Using this result, we can invert \eq{sd2} to obtain

\be \Phi^{(0;0)}_{\b\c\a_1\dots\a_{2s-2}}=\ft18(\C^{\m\n})_{\b\c}
\Phi^{(0;0)}_{\m\n\,\a_1\dots\a_{2s-2}}\ . \la{sd4}\ee
The total symmetry of the left hand side of \eq{sd4} is ensured by
the identity \eq{sd3}. Substituting the trace part $\Phi^{(1;1)}$
in \eq{sd1} by the expression \eq{tp} gives

\bea \nabla_\m \Phi^{(0;0)}_{\n\r\,\a_1\dots\a_{2s-2}}&=&
\ft12(\C_{\n\r})^{\b\c}(\C_\m)^{\d\e}\Phi^{(1;0)}_{\b\c\d\e\a_1\dots
\a_{2s-2}}\la{sd5}\\[15pt]
&&+ \ft{Y}{2(s+2)}\left[
(\C_{\m\n\r})^{\b\c}\Phi^{(0;0)}_{\b\c\a_1\dots\a_{2s-2}}+
(s-1)(\C_\m)_{(\a_1}{}^\b\Phi^{(0;0)}_{\n\r\,\a_2\dots\a_{2s-2})\b}\right]\
.\quad\phantom{q}\nn\eea
In computing the curl and divergence of
$\Phi_{\m\n\,\a_1\dots\a_{2s-2}}$ the contributions from
$\Phi^{(1;0)}$ vanish due to the Fierz identities \eq{fi2} and
\eq{fi4}. It follows that\footnote{Our conventions are
$[\nabla_\m,\nabla_\n]\psi_{a\,\a}=R_{\m\n,a}{}^b\psi_{b\,\a}
-\ft12(\C_{\m\n})_\a{}^\b\psi_{a\,\b}$ and
$\e^{abcde}=i\C^{abcde}$.}:

\bea \nabla_{[\m} \Phi^{(0;0)}_{\n\r]\,\a_1\dots\a_{2s-2}} &=&
-\ft{iY}{12}\e_{\m\n\r}{}^{ab}\Phi^{(0;0)}_{ab\,\a_1\dots\a_{2s-2}}\
, \la{sd6} \\[15pt]
\nabla^\m \Phi^{(0;0)}_{\m\n\,\a_1\dots\a_{2s-2}} &=& 0
,\la{sd7}\eea
where we have used the Fierz identity \eq{fi3} in obtaining
\eq{sd6}. For $Y\neq 0$ the divergence equation \eq{sd7} follows
from the curl equation \eq{sd6}. Taking the divergence of
\eq{sd6} and using \eq{sd7} gives (for all $Y$)

\be
(\nabla^\r\nabla_\r+2s+4-\ft{Y^2}4)\Phi^{(0;0)}_{\m\n\,\a_1\dots\a_{2s-2}}=0\
.\ee
We find (using $2(j_L-j_R)=\D=-Y$ in \eq{ha2}) that the lowest energy
of $\Phi^{(0;0)}_{\m\n\,\a_1\dots\a_{2s-2}}$ is given by $E_0=s+2$
in accordance with \eq{e0}.

The equations \eq{sd6} and \eq{sd7} describe:

\begin{itemize}

\item The physical field equations for the underlined states shown in Table
\ref{tf} (all of which have $|Y|>1$). These are only realized as
two-form potentials in the scalar master field, i.e. they have no
duals in the master gauge field. In particular, at level $\ell=0$
we find the self-duality equation for the $6$-plet of two-form
potentials arising in the supergravity theory.

\item The physical field equations for the remaining spin $s\geq 2$ states
in Table \ref{tf} (all of which have $|Y|>1$). These are realized
as two-form potentials in the scalar master field, and their
field equations take the form of generalized higher spin
self-duality equations in five dimensions. As we shall see below,
these two-form potentials have dual potentials in the master gauge
field.

\item The consistency equations satisfied by the spin $s\geq 1$ Weyl
tensors listed in Table \ref{td} (all of which have $|Y|\leq 1$).
These are non-vanishing curvatures of physical potentials in the
master gauge field.

\end{itemize}


\subsection{The Master Curvature Constraint}


For $|\D|\leq 1$ the set of spin $s$ gauge fields and their
curvature constraints are equivalent to the those used in
\cite{vas1,vas2} to describe a physical spin $s$ field. Those for
$|\D|>1$ are of new type. As we shall see, they are dualized into
the physical two-form potentials with $|Y|>1$ found in $\Phi$ in
the previous section.


\subsubsection*{The Yang-Mills Equations ($s=1$, $\D=0$)}


For spin $s=1$ the curvature constraint \eq{feph} takes the form
$F_{\m\n}=\ft18\Phi^{(0;0)}_{\m\n}$. Identifying
$\Phi^{(0;0)}_{\m\n}$ with the field strength for the spin $s=1$
gauge fields, i.e. the $15$-plet at level $\ell=0$ and the
$(15+1)$-plet at level $\ell=1$, the linearized Yang-Mills
equations follow from \eq{sd7}, while \eq{sd6} becomes the Bianchi
identity (as $Y=0$).


\subsubsection*{The Field Equations for $s\geq \ft32$, $\D=0,\pm 1$ }


For $|\D|\leq 1$ the curvature constraint \eq{feph} decomposes
into physical field equations, generalized torsion equations
(except for spin $s=\ft32$, since there are no auxiliary spin
$s=\ft32$ gauge fields), and identities that give the components
of the curvature which are non-vanishing on-shell, which by
definition are the generalized Weyl tensors.

For $\D=0$ the physical gauge fields are \cite{us,vas1}

\be \ba{l} \D=0\\s=2,3,\dots \ea \ :\quad
A^{(s-1;0)}_{\m\,\a_1\dots\a_{s-1},\b_1\dots
\b_{s-1}}=(\C^{a_1})_{\a_1\b_1}\cdots
(\C^{a_{s-1}})_{\a_{s-1}\b_{s-1}}A_{\m,a_1\dots a_{s-1}}\
,\la{phys}\ee
where $a_1\dots a_{s-1}$ are symmetric and traceless, which
implies that $A^{(s-1;0)}_{(a_1,a_2\dots a_s)}$ is symmetric and
doubly traceless. The $SU(4)$ content is given in the $\D=0$
sector of Table \ref{td}. The $(s-1;0)$ component of \eq{feph} is
a generalized torsion equation which yields the generalized spin
$s$ Lorentz connection
$A^{(s-2;0)}_{\m\,\a_1\dots\a_{s-1},\b_1\dots\b_{s-1}}$ in terms
of one derivative of the physical gauge field. The $(s-2;0)$
component of \eq{feph} contains the physical field equation,
which is of second order and describe a massless spin $s$ field
with lowest energy $E_0=s+2$. It also contains components from
which one can solve for the auxiliary spin $s$ gauge field
$A^{(s-3;0)}$. The $(s-3;0)$, \dots, $(1;0)$ components of
\eq{feph} yield the auxiliary spin $s$ gauge fields
$A^{(s-4;0)}$, \dots, $A^{(0;0)}$, respectively, in terms of
derivatives of the physical gauge field. Finally, the $(0;0)$
component of \eq{feph} sets the non-vanishing component of the
spin $s$ curvature equal to $\Phi^{(0;0)}_{\a_1\dots\a_{2s}}$.
This generalized spin $s$ Weyl tensor is thus given by $s$
derivatives of the physical gauge field, and obeys the equations
\eq{sd6} and \eq{sd7} derived above for $Y=0$.

For $|\D|=1$ the physical gauge fields are \cite{us,vas2,alk}

\be \ba{l} \D= 1\\ s=\ft32,\ft52,...\ea \ :\quad
A^{(s-\fft32;0)}_{\m\,\a_1\dots\a_{s-\fft32}\c,\b_1\dots
\b_{s-\fft32}}=(\C^{a_1})_{\a_1\b_1}\cdots
(\C^{a_{s-\fft32}})_{\a_{s-\fft32}\b_{s-\fft32}}\psi_{\m,a_1\dots
a_{s-\fft32}\,\c}\ ,\la{phys1}\ee
where $a_1\dots a_{s-\ft32}$ are symmetric and $\psi_{\m,a_1\dots
a_{s-\fft32}\,\c}$ is $\C$-traceless ($s\geq \ft52$): $$
(\C^b)_\c{}^\d\psi_{\m,b a_1\dots a_{s-\fft52}\d}=0\ .$$ The
$\C$-tracelessness implies $SO(4,1)$-tracelessness
($s\geq\ft72$). The (complex) dimension of the tangent space irrep
carried by \eq{phys1} is $\ft23(s+\ft32)(s+\ft12)(s-\ft12)$. The
$SU(4)$ content is given in Table \ref{td} for $|\D|=1$. The
$(s-\ft32;0)$ component of \eq{feph} contains the first order
field equation, which gives $E_0=s+2$. For $s\geq \ft52$ it also
yields the auxiliary spin $s$ gauge field
$A^{(s-\fft52;0)}_{\m\,\a_1\dots\a_{s-\fft32}\c,\b_1\dots\b_{s-\fft32}}$
in terms of one derivative of the physical gauge field. The
remaining components of \eq{feph} yield the auxiliary spin $s$
gauge fields $A^{(s-\fft72;0)}$, \dots, $A^{(0;0)}$ and the
generalized Weyl tensor $\Phi^{(0;0)}_{\a_1\dots\a_{2s}}$ in
terms of derivatives of the physical gauge field. The Weyl tensor
satisfies \eq{sd6} and \eq{sd7} for $Y=-\D=\pm 1$.

To summarize, the gauge fields with  $|\D\leq 1$, where $\D$ is
given by \eq{delta}, contain bosonic gauge fields with spin
$s=1,2,\dots$ that have tangent space multi-spinor indices in
one-to-one correspondence \cite{us,5dv1} with two-row $SO(4,1)$
Young tableaux with first row containing $s-1$ boxes and the
second row containing $t$ boxes where $0\leq t\leq s-1$, and
fermionic gauge fields with spin $s=\ft32,\ft52,\dots$
corresponding to tensor-spinors with $SO(4,1)$ indices in the
Young tableaux with first row containing $s-\ft32$ boxes and
second row containing $t$ boxes with $0\leq t\leq s-\ft32$. This
set admits well-known curvature constraints in the tensor basis
\cite{vas1,vas2}. In this paper we have cast them into the
multi-spinor basis (this was first done in the bosonic case in
\cite{us}). For $s\geq \ft32$ the curvature constraints give rise
to physical field equations. For $s=1$ the (linearized)
Yang-Mills equation follows from the master scalar constraint.
The physical field equations are obeyed by the subset of the
gauge fields with $t=0$, given in \eq{phys} and \eq{phys1}. The
remaining ones ($s\geq2$) are auxiliary.


\subsubsection*{The Equations for $|\D| >1$, $s\geq 2$ and Their Dualization}


Let us begin by examining the case of $s=2$ and $|\D|=2$. The only
gauge field present is $A_{\m\,\a\b}$ (the $16_2$-plet listed in
Table \ref{te}). The curvature constraint \eq{feph1} for
$A^{(0;0)}_{\m\,\a\b}$ reads:

\be F_{\m\n\,\a\b} :=
2\nabla_{[\m}A_{\n]\,\a\b}+2(\C_{[\m})_{(\a}{}^\e
A_{\n]\,\b)\e}=\ft18 \Phi^{(0;0)}_{\m\n\,\a\b} ,\la{d4}\ee
where $\Phi^{(0;0)}_{\m\n\,\a\b}$ is the physical two-form
potential obeying the self-duality equations \eq{sd6}. Using
\eq{sd3} and imposing the gauge condition

$$ (\C^\m)_\a{}^\c A_{\m\,\c\b}=0\ ,$$
we find that $A_{\m\,\a\b}$ satisfies the linear field equation

\be (\C^\n)_\a{}^\c\nabla_\n A_{\m\,\c\b}+2A_{\m\,\a\b}=0\
.\la{d5}\ee
Differentiating this\footnote{More generally, a curvature
constraint of the form
$$F_{\m\n\,\a_1\dots\a_m} :=
2\nabla_{[\m}A_{\n]\,\a_1\dots\a_m}+m(\C_{[\m})_{(\a_1}{}^\b
A_{\n]\,\a_2\dots\a_m)\b} =\ft18
\Phi^{(0;0)}_{\m\n\,\a_1\dots\a_m}$$ leads to the first order
equation $$(\C^\n)_{\b}{}^\c\nabla_\n
A_{\m\,\c\a_1\dots\a_{m-1}}+\ft{m+2}2
A_{\m\,\b\a_1\dots\a_{m-1}}=0\ ,$$ in the gauge $\C^\m A_\m=0$.
Differentiating this using $(\backslash \!\!\!\!\!\nabla)^2 A_\m
=(\nabla^2+5+m)A_\m$ one finds the lowest energy $E_0=\ft12
m+2=s+2$. The above constraint arises for $m=2,3$ in gauging
$hs(2,2|4)$ (due to the condition \eq{alge1} there is no gauge
field with $|\D|=4$ and spin $s=3$). Cases with $m>3$ are
expected to arise in gauging enlarged versions of $hs(2,2|4)$.}
one finds the lowest energy $E_0=4$. Thus, interestingly enough,
at the linearized level the spin $s=2$ and $|\D|=2$ state with
$(j_L,j_R)=(\ft32,\ft12)$ has dual formulations in terms of either
a one-form or a two-form potential. An analogous result is valid
for $s=\ft52$ and $|\D|=3$, where the physical gauge field
$A_{\m\,\a\b\c}$ obeys a first order equation with energy
$E_0=\ft92$.

For $s=3$ there is no $|\D|=4$ gauge field (which would have been
physical at the linearized level). The $s=3$ and $|\D|=2$ the
gauge fields are $A^{(0;0)}_{\m\,\a\b\c,\d}$ and
$A^{(1;0)}_{\m\,\a\b\c,\d}$ (the level $\ell=1$ and $\ell=2$
$6_2$-plets listed in in Table \ref{te}). Their curvature
constraint is given by

\be
F_{\m\n\,\a\b\c,\d}:=2\nabla_{[\m}A_{\n]\,\a\b\c,\d}+3(\C_{[\m})_{(\a}{}^\e
A_{\n]\,\b\c)\e,\d}-(\C_{[\m})_\d{}^\e A_{\n]\,\b\c\d,\e}=\ft18
\Phi^{(0;0)}_{\m\n\,\a\b\c\d} .\la{d1}\ee
This constraint decomposes into $(1;0)$, $(1;1)$ and $(0;0)$
parts. The $(1;1)$ part is obeyed identically. The $(1;0)$ part
can be used to solve for $A^{(0;0)}$ in terms of $A^{(1;0)}$:

\be A^{(0;0)}_{\m\,\a\b\c,\d}= \ft1{16}(\C^{ab})_{(\a\b}\left(
2\O_{\m a,b}-\O_{ab,\m} \right)_{\c\d)}\ ,\ee
where

\be \O_{ab,c\,\a\b} = 2(\C_c)^{\c\d} \left( \nabla_{[a}
A^{(1;0)}_{b]\,\a\b\c,\d} +(\C_{[a})_{(\a}{}^\e
A^{(1;0)}_{b]\b)\e\c,\d} \right)\ .\ee
Finally, the $(0;0)$ part yields the curvature of $A^{(0;0)}$ in
terms of $\Phi^{(0;0)}_{\m\n\,\a\b\c\d}$. Thus, one can solve
locally for $A^{(0;0)}$ and $\Phi^{(0;0)}$ in terms of
derivatives of $A^{(1;0)}$ {\it without} going on-shell. By taking
the $\C$-trace of the $(0;0)$ component of \eq{d1} and using
\eq{sd3}, one obtains a second order field equation for
$A^{(1;0)}$. This equation involves rotations of spinor indices,
however, which implies that it is effectively a higher derivative
equation. Thus the physical spin $s=3$ fields with $|Y|=2$
obeying proper physical field equations are the two-form
potentials found in $\Phi$. The curvature constraint should
therefore be interpreted as a duality relation between the
two-form and one-form fields (rather than a field equation for
the latter).

The analysis of the spin $s=3$ and $|\D|=2$ cases generalizes to
$s>3$ and $|\D|>1$. Thus the curvature constraint \eq{feph}
(except the $(0;0)$ part) can be used to solve for auxiliary
gauge fields with $|\D|>1$ in terms of derivatives of independent
gauge fields without going on-shell. The remaining independent
gauge fields are $A^{(s-2;0)}$ ($s=2,3,...$) and
$A^{(s-\fft32;0)}$ ($s=\ft52,\ft72,...$). The remaining $(0;0)$
part of the constraint, which involves $s-1$ derivatives of
$A^{(s-2;0)}$ ($s=2,3,...$) or $s-\ft12$ derivatives of
$A^{(s-\fft32;0)}$ ($s=\ft52,\ft72,...$), dualizes the
independent gauge fields to corresponding higher spin two-form
potentials in $\Phi^{(0;0)}$.

Thus, the independent gauge fields in the $|\D|>1$ sector are:

\bea \ba{l} \D=2\\
s=2,3,\dots\ea &:&
A^{(s-2;0)}_{\m\,\a_1\dots\a_s,\b_1\dots\b_{s-2}} =
(\C^{a_1})_{\a_1\b_1}\cdots
(\C^{a_{s-2}})_{\a_{s-2}\b_{s-2}}(\C^{bc})_{\a_{s-1}\a_s}\nn\\&&\phantom{
A^{(s-2;0)}_{\m\,\a_1\dots\a_s,\b_1\dots\b_{s-2}} =}\quad\times
A_{\m,bc,a_1\dots a_{s-2}}\ ,\la{ind1}\\[15pt]
\ba{l} \D=3\\
s=\ft52,\ft72,\dots\ea &:&
A^{(s-\fft52;0)}_{\m\,\a_1\dots\a_{s-\fft12}\c,\b_1\dots\b_{s-\fft52}}
= (\C^{a_1})_{\a_1\b_1}\cdots
(\C^{a_{s-\fft52}})_{\a_{s-\fft52}\b_{s-\fft52}}
(\C^{bc})_{\a_{s-\fft32}\a_{s-\fft12}} \nn\\&&
\phantom{A^{(s-\fft52;0)}_{\m\,\a_1\dots\a_{s-\fft12}\c,\b_1\dots\b_{s-\fft52}}
= }\quad\times
\psi_{\m,bc,a_1\dots a_{s-\fft52}\c}\
,\la{ind2}\\[15pt]
\ba{l} \D=4\\ s=4,5,\dots\ea &:&
A^{(s-3;0)}_{\m\,\a_1\dots\a_{s+1},\b_1\dots\b_{s-3}} =
(\C^{a_1})_{\a_1\b_1}\cdots
(\C^{a_{s-3}})_{\a_{s-3}\b_{s-3}}(\C^{b_1c_1})_{\a_{s-2}\a_{s-1}}
(\C^{b_2c_2})_{\a_{s}\a_{s+1}}\nn\\&&
\phantom{A^{(s-3;0)}_{\m\,\a_1\dots\a_{s+1},\b_1\dots\b_{s-3}}
=}\quad \times A_{\m,b_1c_1,b_2c_2,a_1\dots a_{s-3}}\ .\la{ind3}\eea
Here the Young tableaux for $bc,a_1\dots a_{s-2}$ ($|\D|=2$) has
$s-1$ boxes in the first row and $1$ box in the second row and
dimension $(s+2)(s+\ft12)(s-1)$; the Young tableaux for
$b_1c_1,b_2c_2,a_1\dots a_{s-3}$ ($|\D|=4$) has $s-1$ boxes in
the first row and $2$ boxes in the second row and dimension
$\ft53 (s+3)(s+\ft12)(s-2)$. The tensor-spinor in the right hand
side of \eq{ind2}, which has (complex) dimension $\ft43
(s+\ft52)(s+\ft12)(s-\ft32)$ obeys suitable $\Gamma$-trace
conditions. All these gauge fields are dualized to two-form
potentials, which obey the physical field equation \eq{sd6}
realizing the $|Y|>1$, $s\geq 2$ sector of the spectrum in Table
\ref{tc}. The spin $s=2$, $\D=2$ gauge field $A_{\m\,\a\b}$ and
spin $s=\ft52$, $\D=3$ gauge field $A_{\m\,\a\b\c}$ also obey
physical field equations, which assume the form \eq{d5} in a
fixed gauge.

To summarize, the gauge fields with $|\D|>1$) are of new type.
All gauge fields in this set have spin $s\geq 2$. For
$s=2,3,\dots$ their tangent space multi-spinor indices are in
one-to-one correspondence with two-row $SO(4,1)$ Young tableaux
with first row containing $s-1$ boxes and the second row
containing $t$ boxes where $0\leq t\leq s-2$. For spin
$s=\ft52,\ft72,\dots$ the $SO(4,1)$ Young tableaux have $s-\ft32$
boxes in the first row and $t$ boxes in the second row with
$0\leq t \leq s-\ft52$. We have identified the linearized
curvature constraints for the gauge fields in this set. As a
result the gauge fields listed in \eqs{ind1}{ind3} remain
independent, while the other gauge fields can be expressed in
terms of derivatives of the independent gauge fields without
imposing any field equations. Higher derivative field equations
for the independent gauge fields (for $s>2$) arises by
$\C$-tracing the remaining curvature component. The
$\C$-traceless part of this component implies a dualization of
the independent one-forms into (higher spin) two-form potentials
contained in the master scalar field. These two-forms obey physical field
equations generalizing the self-duality equation in five
dimensions satisfied by the two-form potential in the gauged
supergravity sector \cite{5d1,5d2}.


\section{Conclusions}


In this paper we have taken the first step towards the
construction of the full interacting theory by identifying the
symmetry group, the full spectrum, the master fields required to
describe the theory and the correct linearized equations of
motion. In particular, we have shown how the linearized field
equations of $D=5,N=8$ gauged supergravity are embedded in the
theory. For example, the correct $AdS_5$ mass splitting among the
42 scalars and the self duality equation for the 6-plet of
two-form potentials are obtained. These results provide
nontrivial checks on some crucial aspects of the theory, such as
the conditions imposed on the master scalar field, as well as the
basic hypothesis that the full spectrum of the theory consists of
all states resulting from the symmetric product of two $d=4, N=4$
Yang-Mills superdoubleton. One curious result is that the full
spectrum exhibited in Table 3 coincides precisely with the
spectrum of higher spin $D=4, N=8$ supergravity \cite{us4} for
all levels above the lowest one! Thus, the $PSU(2,2|4)$
supermultiplets in levels $\ell=1,2,...,\infty$ of the
five-dimensional theory listed in Table 3 seem to be also the
supermultiplets of the $AdS_4$ superalgebra $OSp(8|4)$. The full
spectra seem to differ only at level $\ell=0$, where the distinct
supergravity multiplets reside.

It is clear that the next step is to construct the full
interacting theory. The existence of certain cubic interactions
for the $|\D\leq 1$ gauge fields \cite{5dv1} indicates that there
exist consistent interactions to all orders, though these
necessarily require the inclusion of the ``matter'' fields
contained in $\Phi$. Experience from four spacetime dimensions,
where the interacting full theory of massless higher spins exists,
suggests that a natural framework for discussing the interactions
is an extended spacetime with extra non-commutative ``z''-space
coordinates \cite{v7} (See \cite{us4} for a detailed study of the
case of higher spin $D=4, N=8$ supergravity). In fact, the
introduction of such a space is straightforward in the present
case as well. Moreover, we expect that the full interacting higher
spin $D=5, N=8$ supergravity equations will be described by
z-extended and suitably ``twisted'' versions of our constraint
equations \eq{feph} and \eq{dphez}, to be  imposed, of course, on
the full curvature two-form and full covariant derivative of the
master zero-form. We also expect the resulting theory to yield a
curvature expansion, just as in the four dimensional case.  It
would be interesting to compare the results of this expansion
with the bulk predictions of the boundary CFT based on higher spin
currents formed out of the $d=4, N=4$ Yang-Mills theory in the
limit of zero t'Hooft coupling and large dimension of the gauge
group.

Another direction, motivated by the question of whether there
exists non-trivial interactions for tensionless opens strings, is
to examine various extensions \cite{5dv2} of the higher spin gauge
group $hs(2,2|4)$, and their relation to (higher spin)
superdoubletons with non-vanishing central charge. For example,
one may restrict the $\t_\y$-condition only to certain values of
$\y$. We expect that the new generators will give rise to gauge
fields that are dual to additional higher spin two-form
potentials. The extensions appears to prevent modding out the
central charge, however, in which case the spectrum will contain
infinitely many massless higher spin fields of each given spin.
Whether such a degeneracy  is natural from the point of view of
string theory is not altogether obvious because the zero tension
limit of Type IIB string theory is intrinsically nonperturbative
from the worldsheet point of view ( as $\a' \ra \infty$) and not
much is known about the space of soliton solutions in this setup
that can be interpreted as physical states in $AdS_5$
\footnote{We thank M. Douglas for illuminating discussions on this
point.}.

An important obstacle in making progress towards establishing a
duality between the free limit of $N=4$ SYM and a higher spin
gauge theory in the bulk is really our lack of knowledge about the
quantization of Type IIB strings in $AdS_5 \times S^5$. Hardly
anything is known about the string states beyond the low energy
supergravity and the Kaluza-Klein modes in this case. It is true
that some scattered and interesting variety of soliton solutions
in Type IIB theory are known but their relevance or fate in the
tensionless limit is not clear.

With this state of affairs, one may hope that the quantization of
tensionless Type IIB string on $AdS_5 \times S^5$, unlike in the case of
flat target, is  more amenable to study. If this approach proves
to be just as difficult as the finite tension case, one avenue
which still remains open and technically within reach, is to study
the $4d$ CFT based on the free limit of $N=4$ SYM (or its higher
spin extension thereof) in detail, e.g. its correlation functions
\cite{su1,su2}, and to try to obtain information about the bulk
theory they point to. In this way, one may not only try to
establish a connection with an interacting higher spin gauge
theory in the bulk, but by establishing the properties of such a
theory (e.g. its spectrum, symmetries and symmetry breaking
mechanisms) one may also get an idea about where to look for the
states of the tensionless Type IIB string.

\vspace{5pt}


\section*{Acknowledgements}


This research project has been supported in part by NSF Grant
PHY-0070964. The work of P.S. is part of the research program of
Stichting voor Fundamenteel Onderzoek der Materie (Stichting FOM).
We thank Feza G{\"u}rsey Institute, Groningen University and
Benasque Center for Science for hospitality. We also thank O.
Aharony, M. Berkooz, M. Douglas, A. Polyakov, K. Skenderis and A.
Zaffaroni for useful discussions on various aspects of
tensionless strings.

\pagebreak

\begin{appendix}

\section{Decomposition of Spectrum into Massless $PSU(2,2|4)$ Irreps}

\def\ket#1{|#1\rangle}

The spectrum $\cS$ of the $hs(2,2|4)$ gauge theory is given by the
symmetric part of the tensor product of two Maxwell multiplets
(superdoubletons with $Z=0$). It is well-known \cite{g1} that this
space decomposes into massless irreps of $PSU(2,2|4)$. In order
to apply this result to the construction of the $hs(2,2|4)$ gauge
theory we need to find by what multiplicity each massless
$PSU(2,2|4)$ multiplet occurs in ${\cal S}$.

To this end, we split the $SU(2,2)\times SU(4)$-covariant
super-oscillator $Z_R\equiv (y_{\underline \a},\th_i)$,
$\underline{\a}=1,\dots,4$; $i=1,\dots,4$, into $SU(2)_L\times
SU(2)'_{L}\times SU(2)_R\times SU(2)'_{R}$-covariant
superoscillators $\x_A$ and $\y^K$ as (see \eq{a8} and \eq{a9}):

\bea
Z_R&\rightarrow& \sqrt{2}(\x_A,\y^K)\ ,\nn\\[15pt]\x_A&=&(\x_a,\x_\a),
\ a=1,2;\ \a=1,2\ ,
\nn\\[15pt]
\y^K&=&(\y^k,\y^\kappa),\ k=1,2;\  \kappa=1,2\ ,
\la{caosc1}
\eea
where $(-1)^{a}=-(-1)^\a=(-1)^k=-(-1)^{\k}=1$. Defining

\be \x^A=(\x_A)^{\dagger}\ ,\quad \eta_K=(\eta^K)^{\dagger}\ ,\la{caosc2}\ee
the oscillator algebra \eq{osc} is equivalent to the following
graded commutation rules:

\be \x_A\star \x^B-(-1)^{AB}\x^B \star \x_A=\delta_A^B\ ,\quad
\eta_K\star \eta^L-(-1)^{KL}\eta^L\star\eta_K=\delta_K^L\ ,\ee
The vacuum state $\ket{0}$ of the oscillator Fock space is defined
by

\be \xi_A\ket{0}=0\ ,\quad \eta_K\ket{0}=0\ .\ee
The $SU(2,2|4)\times U(1)_Y$ generators decompose
into

\be \Lambda_{AK} :=\xi_A\eta_K\ ,\quad \xi^A\xi_B\ ,\quad
\y^K\y_L\ee
and hermitian conjugates. The energy operator is given by

\be E=\ft12(\x^a\x_a+\y^k\y_k+2)\ ,\ee
and the $U(1)$-charges are given by

\be Z=\ft12(\x^a\x_a+\x^\a\x_\a-\y^k\y_k-\y^\k\y_\k)\ ,\quad
Y=\x^\a\x_\a-\y^\k\y_\k\ .\ee
The lowest weight states of the maximal bosonic subalgebra
$B=SU(2,2)\times SU(4)\times U(1)_Z\subset SU(2,2|4)$ are
annihilated by the bosonic energy-lowering operators $\L_{ak}$
and $\L_{\a\k}$ and labeled by the energy $E_0$, the
$SU(2)_L\times SU(2)_R$ spins $(j_L,j_R)$, an $SU(4)$ irrep $R$
and a central $U(1)_Z$ charge $Z$. They also carry a $U(1)_Y$
charge $Y$.

The split \eq{caosc1} defines a realization of the maximal
compact supergroup $SU(2|2)_L\times SU(2|2)_R\subset SU(2,2|4)$.
The bosonic subalgebra $B'=SU(2)_L\times SU(2)'_{L}\times
SU(2)_R\times SU(2)'_{R}\subset SU(2|2)_L\times SU(2|2)_R$ has
the generators:

\be L^a_b=\x^a\x_b-\ft12 \delta^a_b\x^c\x_c\ ,\quad
L^{\prime}{}^\a_\b=\x^\a\x_\b-\ft12\d^\a_\b \x^\c\x_\c\ ,
\la{a8}\\[15pt]
\ee
\be
R^k_l=\y^k\y_l-\ft12 \d^k_l \y^m\y_m\ ,\quad
R^\prime{}^\k_\l=\y^\k\y_\l-\ft12\d^\k_\l \y^\m\y_\m\ .
\la{a9}
\ee
The Clifford vacuum, $\ket{\O}$, of an $SU(2,2)$ supermultiplet
is a lowest weight state annihilated by bosonic as well as
fermionic energy-lowering operators, that is

\be \L_{AK}\ket{\O}=0\ .\la{lwsc}\ee
This ground state consists of a set of lowest weight states of
the bosonic subalgebra $B$ forming a supermultiplet of
$SU(2|2)_L\times SU(2|2)_R$. The states in $\ket{0}$ can
therefore alternatively be labeled by their $B'$ highest weights
$(j_L,j_R;j'_L,j'_R)$ and $U(1)_Y$ charge. Parametrizing
such a state with central charge $Z=\ft12(m-n)$ as

\be \ket{\Omega)}=\psi_{A_1\dots A_m;K_1\dots K_n}\x^{A_1}\cdots
\x^{A_m}\eta^{K_1}\cdots \eta^{K_n} \ket{0} \ ,\ee
and imposing \eq{lwsc} one finds $mn\psi_{A A_1\dots A_{m-1}; K
K_1\dots K_{n-1}}=0$. Thus the Fock space of the superoscillators
decomposes into superdoubletons with with Clifford vacua
$\ket{\O}=\x^{A_1}\cdots\x^{A_{2j}}\ket{0}$ and
$\ket{\O}=\y^{K_1}\cdots\y^{A_{2j}}\ket{0}$,
$j=0,\ft12,1,\ft32,...$, with central charge $Z=\pm j$,
respectively. Acting on them with the remaining supercharges
$\x^\a\y^k$ and $\x^a\y^\k$ gives the supermultiplets \cite{g1}
listed in Tables \ref{ta} and \ref{tb}.

In order to decompose the product of two doubletons we describe
the tensor product by introducing two flavors of oscillators
labeled by an index $r=1,2$ as follows:

\be [\x(r)_A,\x^B(s)]=\delta_{rs}\d_A^B\ ,\quad
[\y(r)_K,\y^L(s)]=\delta_{rs}\d_K^L\ ,\quad r,s=1,2\ .\ee
An $SU(2,2|4)\times U(1)_Y$ generator, $\L$ say, is represented by
$\L=\sum_{r=1,2}\L(r)$. In particular
$$E=2+\ft12\sum_{r=1,2}(\x^a(r)\x_a(r)+\y^k(r)\y_k(r))\ ,$$which shows that
irreps of the two-fold tensor product are massless. We are interested in the
product ${\cal D}\otimes{\cal D}$ of two $Z=0$ weight spaces ${\cal D}$,
each of which consist of the states:

\be {\cal D}=\ket{0}\oplus \left(\x^A\y^K\ket{0}\right)\oplus
\left( \x^{A}\x^{B}\y^{K}\y^{L}\right)\ket{0}\oplus\dots\ .\ee
Parametrizing a ground state in ${\cal D}\otimes{\cal D}$ as ($n\in {\sf Z}$):

\bea \ket{\O}=\sum_{m=0}^n &&\psi^{(m)}_{A_1\dots A_m,B_1\dots
B_{n-m}; K_1\dots K_m,L_1\dots L_{n-m}}\\&&\x^{A_1}(1)\cdots
\x^{A_m}(1)\x^{B_1}(2)\cdots \x^{B_{n-m}}(2) \y^{K_1}(1)\cdots
\y^{K_m}(1)\y^{L_1}(2)\cdots \y^{L_{n-m}}(2)\ket{0}\ .\nn\eea
and imposing $(\L_{CM}(1)+\L_{CM}(2))\ket{\O}=0$ we find the
conditions

\bea \psi^{(1)}_{C,B_1\cdots B_{n-1};M,L_1\cdots
L_{n-1}}+n^2\psi^{(0)}_{CB_1\cdots B_{n-1};ML_1\cdots
L_{n-1}}&=&0\ ,\nn\\[15pt]
4\psi^{(2)}_{CA_1,B_1\cdots B_{n-2};MK_1,L_1\cdots
L_{n-2}}+(-1)^{CA_1+MK_1}(n-1)^2\psi^{(1)}_{A_1,CB_1\cdots
B_{n-2};K_1,ML_1\cdots
L_{n-2}}&=&0\ ,\nn\\[15pt]
&\vdots&\nn\\
n^2\psi^{(n)}_{CA_1\cdots A_{n-1};MK_1\cdots
K_{n-1}}+(-1)^{C(A_1+\cdots+A_{n-1})+M(K_1+\cdots+K_{n-1})}\psi^{(n-1)}_{
A_1\cdots A_{n-1},C;K_1\cdots
K_{n-1},M}&=&0\ .\nn\\[15pt] \eea
It follows that for each $n$ there is a unique ground state. We
find that the tensor product ${\cal D}\otimes {\cal D}$
decomposes into massless supermultiplets ${\cal D}_{s_{\rm max}}$
(with maximal spin $s_{\rm max}$) labeled by Clifford vacua
$\ket{\O_{s_{\rm max}}}$ as follows:

\bea &&\ket{\Omega_2}=\ket{0}\ ,\\[15pt]
&&\ket{\O_3}=\left[\x^A(1)\y^K(1)-\x^A(2)\y^K(2)\right]\ket{0}\ ,\nn\\[15pt]
&&\ket{\O_4}=\left[\x^{A_1}(1)\x^{A_2}(1)\y^{K_1}(1)\y^{K_2}(1)-
4\x^{[A_1}(1)\x^{A_2]}(2)\y^{[K_1}(1)\y^{K_2]}(2)+
\x^{A_1}(2)\x^{A_2}(2)\y^{K_1}(2)\y^{K_2}(2)\right]\ket{0}\ ,\nn\\
&&\vdots\nn\\
&&\ket{\O_{n+2}}=\nn\\&&\sum_{m=0}^n(-1)^m{n\choose m}^2
\x^{[A_1}(1)\cdots\x^{A_m}(1)\x^{A_{m+1}}(2)\cdots\x^{A_n]}(2)
\y^{[K_1}(1)\cdots\y^{K_m}(1)\y^{K_{m+1}}(2)\cdots\y^{K_n]}(2)
\ket{0}\ ,\nn\\
&&\vdots \nn\eea
where $[A_1\dots A_n]$ denotes graded symmetrization. The
symmetric and anti-symmetric parts of the tensor product are
given by

\bea \left({\cal D}\otimes{\cal D}\right)_S&=&
\sum_{\ell=0,1,\dots} {\cal D}_{s_{\rm max}=2\ell+2}\ ,\\[15pt]
\left({\cal D}\otimes{\cal D}\right)_A&=& \sum_{\ell=0,1,\dots}
{\cal D}_{s_{\rm max}=2\ell+3}\ ,\eea
where $\ell$ is a level index. In the symmetric product, the level
$\ell=0$ multiplet is the $D=5$, $N=8$ supergravity multiplet;
the level $\ell\geq 1$ multiplets have spin range $4$ and consist
of $256+256$ states. The $SU(2,2)\times SU(4)\times U(1)_Y$
content of these multiplets \cite{g1} is listed in Table \ref{tc}.


\section{Harmonic Analysis on $AdS_5$}


To determine the $SO(4,2)$ content of the spectrum, we shall
follow the technique used in \cite{es1,es2} which is based on the
analytic continuation of $AdS_5$ to the five-sphere, and
consequently the group $SO(4,2)$ to $SO(6)$. The Casimir
eigenvalues for an $SO(4,2)$ representation $D(j_L,j_R;E_0)$ and
an $SO(6)$ representation with highest weight labels
$(n_1,n_2,n_3)$ $(n_1\geq n_2\geq |n_3|$) are

\bea C_2[SO(4,2)] &=& E_0(E_0-4) +2j_L(j_L+1)+2j_R(j_R+1) \nn\w2 \
C_2[SO(6)] &=& n_1(n_1+4) + n_2(n_2+2) + n_3^2 \ ,\la{ha1} \eea
where the continuation requires the identification:

\be \nabla^2|_{AdS_5}\rightarrow -\nabla^2|_{S^5}\ ,\quad
E_0=-n_1\ ,\quad n_2=j_L+j_R\ ,\quad n_3=j_L-j_R\ .\la{ha2}\ee
The $SO(6)$ Casimir is related to the Laplacian acting on a
tensor $T$ on $S^5$ in an irrep $R$ of $SO(5)$, which we expand as

\be T_{\a_1\dots\a_{2s}}(x)=\sum_{\tiny \ba{c}(n_1,n_2,n_3)\\p\ea}
T_p^{(n_1,n_2,n_3)}
D^{(n_1,n_2,n_3)}_{\a_1\dots\a_{2s},p}(L_x^{-1})\ ,\la{ha3}\ee
by the following formula

\be -\nabla^2|_{S^5}D^{(n_1,n_2,n_3)}_{\a_1\dots\a_{2s},p} =\left(
C_2[SO(6)]-C_2[SO(5)]\right)D^{(n_1,n_2,n_3)}_{\a_1\dots\a_{2s},p}\
.\la{ha4}\ee
In \eq{ha3}, $L_x$ is a coset representative of a point $x\in
S^5$ and $(n_1,n_2,n_3)$ label all $SO(6)$ representations
containing $R$, namely those which satisfy the embedding $n_1\geq
m_1\geq n_2\geq m_2\geq |n_3|$ where $(m_1,m_2)$ ($m_1\geq m_2\geq
0$) are the highest weight labels of $R$. The $SO(5)$ Casimir is
given by

\be C_2[SO(5)] = m_1(m_1+3) +m_2(m_2+1)\ .\ee
Using the notation introduced in \eq{rt}, the $SO(5)$ highest
weight labels of $\Phi^{(r;0)}_{\a_1\dots\a_{2s}}$ ($0\leq r \leq
s$) are

\be m_1=s\ ,\quad m_2=s-r\ .\la{hwrs}\ee
The dimension of this representation is given by

\be d_{m_1,m_2}=\ft23 (m_1+\ft32)(m_2+\ft12)(m_1+m_2+2)(m_1-m_2+1)\ .\ee
In case of integer spin $s$, the irrep $R$ corresponds to an
$SO(4,1)$ Young tableaux with $m_1$ boxes in the first row and
$m_2$ boxes in the second row.

\end{appendix}

\pagebreak


\end{document}